\documentclass[prc,a4paper,floatfix,nofootinbib]{revtex4}
\usepackage{amsmath,amsfonts,amssymb,graphicx}

\newcommand{\capindex}[1]{\tiny{#1}}

\newcommand{\BCS}{\text{\capindex{BCS}}}
\newcommand{\D}{\text{\capindex{D}}}
\newcommand{\dd}{d}
\newcommand{\e}{e}
\newcommand{\eps}{\varepsilon}
\newcommand{\F}{\text{\capindex{F}}}
\newcommand{\fb}{f_{\text{\capindex{B}}}}
\newcommand{\ff}{f_{\text{\capindex{F}}}}
\newcommand{\fm}{\text{fm}}
\newcommand{\fmi}{\text{fm}^{-1}}
\newcommand{\HF}{\text{\capindex{HF}}}
\newcommand{\ii}{i}
\newcommand{\kk}{\textbf{\text{k}}}
\newcommand{\lad}{\text{\capindex{E}}}
\newcommand{\lowk}{\text{low-}k}
\newcommand{\MeV}{\text{MeV}}
\newcommand{\NSR}{\text{\capindex{NSR}}}
\newcommand{\pp}{\mathbf{p}}
\newcommand{\principalvalue}{\mathcal{P}\!}

\newcommand{\sHF}{\text{s-\capindex{HF}}}
\newcommand{\stot}{\text{s-tot}}
\newcommand{\transpose}{\intercal}

\newcommand{\Eq}[1]{Eq.~\eqref{#1}}
\newcommand{\Fig}[1]{Fig.~\ref{#1}}
\newcommand{\Figs}[1]{Fig.~\ref{#1}}
\newcommand{\Reference}[1]{Ref.~\cite{#1}}
\newcommand{\Refs}[1]{Refs.~\cite{#1}}
\newcommand{\Sec}[1]{Sec.~\ref{#1}}

\DeclareMathOperator{\Real}{Re}
\DeclareMathOperator{\Imag}{Im}
\renewcommand{\Re}{\Real}
\renewcommand{\Im}{\Imag}

\begin{document}
%%%%%%%%%%%%%%%%%%%%%%%%%%%%%%%%%%%%%%%%%%
\title{BCS-BEC Crossover Effects and Pseudogap in Neutron Matter}
\author{David Durel}\email{david.durel@ijclab.in2p3.fr}
\author{Michael Urban}\email{michael.urban@ijclab.in2p3.fr}
\affiliation{Universit\'e Paris-Saclay, CNRS/IN2P3, IJCLab,
  91405 Orsay cedex, France}
\begin{abstract}
  Due to the large neutron-neutron scattering length, dilute
  neutron matter resembles the unitary Fermi gas, which lies half-way
  in the crossover from the BCS phase of weakly coupled Cooper
  pairs to the Bose-Einstein condensate of dimers. We discuss
  crossover effects in analogy with the T-matrix theory used in the
  physics of ultracold atoms, which we generalize to the case of a
  non-separable finite-range interaction. A problem of the standard
  Nozi\`eres--Schmitt-Rink approach and different ways to solve it are
  discussed. It is shown that in the strong-coupling regime, the
  spectral function exhibits a pseudo-gap at temperatures above the
  critical temperature $T_c$. The effect of the correlated density on
  the density dependence of $T_c$ is found to be rather weak, but a
  possibly important effect due to the reduced quasiparticle weight is
  identified.
\end{abstract}

\maketitle

%%%%%%%%%%%%%%%%%%%%%%%%%%%%%%%%%%%%%%%%%%%%%%%%%%%%%%%%%%%%%%%%%%%%%%%%%%%%%%%
\section{Introduction}
%%%%%%%%%%%%%%%%%%%%%%%%%%%%%%%%%%%%%%%%%%%%%%%%%%%%%%%%%%%%%%%%%%%%%%%%%%%%%%%
The inner crust of neutron stars is characterized by the presence of a
dilute gas of neutrons in between much denser clusters
\cite{Chamel2008}. The superfluidity of this gas is of decisive
importance, e.g., to explain pulsar glitches and to determine the
thermal evolution of neutron stars. Nevertheless, in spite of a long
effort and the existence of Quantum Monte-Carlo (QMC) calculations of
the gap at low densities \cite{Abe2009,Gezerlis2010}, there remain
large uncertainties about the density dependence of the pairing gap
and the critical temperature of neutron matter.

In reality, the neutron gas in the inner crust is not uniform because
of the nuclear clusters. However, if the pairing is strong enough and
the Cooper-pair size (coherence length) small enough, i.e., much
smaller than the distance between clusters, it should be possible to
treat the neutron gas approximately as uniform matter, except near the
cluster surface \cite{Okihashi2020}. Furthermore, in the cases where
the local-density approximation is not valid, the calculations rely
generally on density-dependent effective interactions, fitted such
that they reproduce, at the mean-field level, the realistic pairing
gaps in uniform matter \cite{Chamel2010,Okihashi2020}. So, also in
this case, the understanding of uniform neutron matter is a necessary
prerequisite to describe superfluidity of the inner crust.

The neutron-neutron ($nn$) interaction has a very large scattering
length $a \simeq -18.7 \pm 0.6\;\fm$~\cite{Gonzales1999} or
$-16.59\pm 1.17\;\fm$~\cite{Babenko2013}) while its effective range is
only $r_{0} = 2.83\pm 0.11\;\fm$~\cite{Babenko2013}. Hence, at low
densities, when the mean distance between the neutrons is between $3$
and $16\;\fm$, the neutron gas resembles a unitary Fermi gas, an
idealized system of spin-$1/2$ Fermions with a contact interaction
having infinite scattering length, which was originally introduced as
a schematic model of dilute neutron matter \cite{Baker1999}.

A few years later, the unitary Fermi gas was experimentally realized
with ultracold trapped gases of alkali atoms, where the interaction
has practically zero range (four orders of magnitude smaller than the
interparticle distance) and the scattering length can be tuned with
the help of a magnetic field from negative to positive values by
passing through infinity at the so-called Feshbach resonance. This
allows one to study not only the unitary limit, but the whole
crossover from BCS pairing with Cooper pairs in the case of weakly
attractive interactions ($a<0$) to the Bose-Einstein condensation
(BEC) of bound dimers in the case $a>0$ \cite{CalvaneseStrinati2018}.
The unitary limit is just a special case in this crossover,
corresponding to the situation where the binding energy of the dimer
in free space tends to zero.

On the BEC side and in the crossover region, it is crucial for a
correct description of the superfluid phase transition that pair
correlations persist at temperatures above the critical temperature
$T_c$. On the BEC side of the crossover, these correlations can easily
be interpreted as dimers which are already formed but not yet
condensed. This implies in particular that the BCS relation $T_c =
0.57 \, \Delta(0)$ between the critical temperature $T_c$ and the
zero-temperature pairing gap $\Delta(0)$ is not fulfilled in this
regime. The simplest theory that correctly interpolates between the
BCS and the BEC limits was proposed by Nozi\`eres and Schmitt-Rink
(NSR) \cite{Nozieres1985}.

In nuclear systems, the interaction is of course fixed. Nevertheless,
a BCS-BEC crossover is expected to happen in symmetric nuclear matter
if one varies the density: at very low density, there will be a BEC of
deuterons, which continuously goes over to a BCS superfluid with
proton-neutron Cooper pairs at higher density \cite{Baldo1995}. The
critical temperature $T_c$ in this crossover was studied in
\cite{Schmidt1990,Stein1995,Jin2010} using an approach similar to the
NSR theory.

The case of neutron matter is somewhat different, because no bound
dineutron state and hence no BEC phase exist. Nevertheless, with
varying density, neutron matter passes from the strongly coupled
regime close to the unitary limit to the weakly coupled BCS
regime. The relevance of BCS-BEC crossover-like physics in dilute
neutron matter was pointed out in \cite{Matsuo2006}. Consequently,
NSR-like corrections of $T_c$ due to pair correlations in the normal
phase were studied in
\cite{Ramanan2013,Ramanan2018,Tajima2019,Ramanan2020}, and in
\cite{Ohashi2020,Inotani2019} the NSR-like approach was also extended
to temperatures below $T_c$.

In previous work following the NSR and related approaches, the
single-particle self-energy and spectral function were usually not
computed. One reason is that in these approaches the density $\rho$ is
often obtained as the derivative of the thermodynamic potential with
respect to the chemical potential $\mu$
\cite{Nozieres1985,SadeMelo1993,Tajima2019,Ohashi2020,Inotani2019}.
Computing the thermodynamic potential in the ladder approximation, one
automatically obtains an expression which is equivalent to truncating
the Dyson equation for the single-particle Green's function at the
first order in the self-energy \cite{CalvaneseStrinati2018}. In this
case, when calculating only the density $\rho$ and not the occupation
numbers $n(k)$, the somewhat difficult computation of the self-energy
can be avoided. However, some problems of the different approaches
show up when one looks at the occupation numbers $n(k)$, and some
interesting aspects of the crossover regime, such as the existence of
a pseudogap, can only be studied by looking at the spectral
function. In this context, let us note that the density of
  states, which is usually considered when discussing the pseudogap,
  is obtained by integrating the spectral function over momentum.

In the present work, we revisit this problem, but now we compute the
self-energy, which gives us access to the occupation numbers and to
the spectral function. In \Sec{sec:formalism}, we recall the T matrix
formalism and discuss in some detail the way we deal with
non-separable interactions. In \Sec{sec:results}, we will show our
results for the spectral function, the pseudogap, the correlated
occupation numbers, and the density dependence of the critical
temperature. We will in particular discuss the differences between the
original NSR approach, its modifications that have been used
previously, and our more complete treatment of the self-energy. Some
open questions are discussed in \Sec{sec:open}, and we conclude in
\Sec{sec:conclusions}.

%%%%%%%%%%%%%%%%%%%%%%%%%%%%%%%%%%%%%%%%%%%%%%%%%%%%%%%%%%%%%%%%%%%%%%%%%%
\section{T-matrix formalism}
\label{sec:formalism}
%%%%%%%%%%%%%%%%%%%%%%%%%%%%%%%%%%%%%%%%%%%%%%%%%%%%%%%%%%%%%%%%%%%%%%%%%%
%%%%%%%%%%%%%%%%%%%%%%%%%%%%%%%%%%%%%%%%%%%%%%%%%%%%%%%%%%%%%%%%%%%%%%%%%%
\subsection{Vertex function}
\begin{figure}
\begin{center}
\includegraphics[width=15cm]{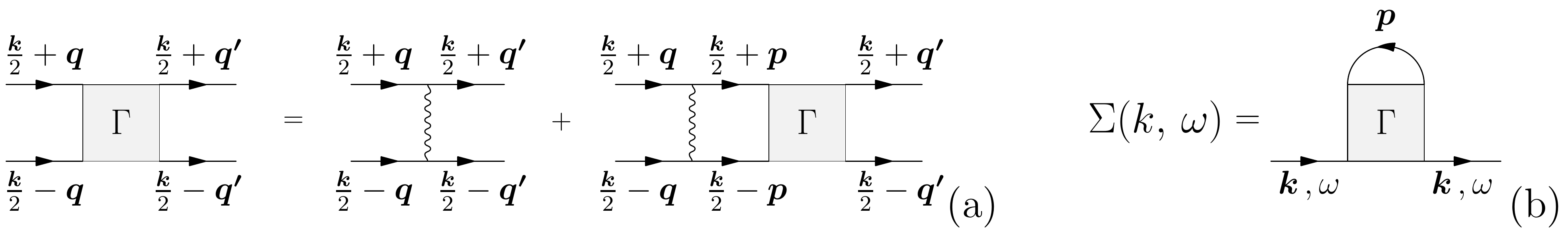}
\caption{Feynman diagrams for (a) the vertex function $\Gamma$ and (b)
  the self-energy $\Sigma$ in ladder approximation.
\label{fig:feynman}}
\end{center}
\end{figure}

%%%%%%%%%%%%%%%%%%%%%%%%%%%%%%%%%%%%%%%%%%%%%%%%%%%%%%%%%%%%%%%%%%%%%%%%%%
The superfluid phase transition is an instability of the normal phase
towards the superfluid one which shows up in the two-particle T matrix
(vertex function) $\Gamma$. We compute $\Gamma$ within the ladder
approximation, i.e., by resumming ladder diagrams in the medium.  In
this formalism, the vertex function for total momentum $k$, in- and
outgoing relative momenta $q$ and $q'$, and total energy of the pair
$\omega$, is the solution of the following Lippmann-Schwinger like
equation corresponding to the Feynman diagrams shown in
\Fig{fig:feynman}(a),
%%%%%%%%%%%%%%%%%%%%%%%%%%%%%%%%%%%%%%%%%%%%%%%%%%%%%%%%%%%%%%%%%%%%%%%%%%%%%%%%
%%%%%%%%%%%%%%%%%%%%%%%%%%%%%%%%%%%%%%%%%%%%%%%%%%%%%%%%%%%%%%%%%%%%%%%%%%%%%%%%
\begin{equation}\label{eq_lip_sch_4}
\Gamma(q, q', k, \omega) = v(q, q')
  + \int \dfrac{\dd^{3} \pp}{(2\pi)^{3}} \, v(q, p) \,
  G_{0}^{(2)}(\pp, \kk, \omega) \, \Gamma(p, q', k, \omega)\,,
\end{equation}
where $v(q, q')$ is the matrix element of the $s$-wave $nn$
interaction. In our calculations, we will use interactions of the
$V_{\lowk}$ type \cite{Bogner2010}. Notice that, in contrast to
nuclear matter with protons or very dense neutron matter
\cite{Holt2010}, the effect of three-body interactions is not
important in dilute neutron matter because the leading contact term is
forbidden by the Pauli principle.

At finite temperature, the two-particle Green's function $G_{0}^{(2)}$
is written within the Matsubara formalism \cite{FetterWalecka} as
\begin{equation}
  G_{0}^{(2)} (\pp, \kk, \ii \Omega_N) = \dfrac{1}{\beta}
  \sum_{n} G_{0}(|\tfrac{\kk}{2} + \pp|, \ii\omega_{n}) \,
  G_{0}(|\tfrac{\kk}{2} - \pp|, \ii \Omega_{N} - \ii\omega_{n} )\,,
\end{equation}
where $ \xi_{k} = \hbar^2 k^{2}/(2m) - \mu $ is the neutron
single-particle energy relative to the chemical potential $\mu$, with
$\hbar$ the reduced Planck constant and $m$ the neutron mass, $\beta =
1/T$ is the inverse of the temperature $T$, $\Omega_N = 2\pi N T$ and
$\omega_n = (2n+1)\pi T$ are bosonic and fermionic Matsubara
frequencies, respectively, and $G_0(k,\ii\omega_n) =
1/(\ii\omega_n-\xi_k)$ is the free single-particle Green's
function. After summation over $\omega_n$ and analytic continuation
to the real $ \omega $ axis, we obtain the retarded two-particle
Green's function
\begin{equation}\label{G02}
  G_{0}^{(2)} (\pp, \kk, \omega) =
  \dfrac{1 - \ff(\xi_{\kk/2 + \pp}) - \ff(\xi_{\kk/2 - \pp})}
        {\omega - \xi_{\kk/2 + \pp} - \xi_{\kk/2 - \pp} + \ii \eps}\,,
\end{equation}
where $\ff(\xi) = 1/(\e^{\beta\xi}+1)$ denotes the Fermi function. The
limit $\eps\to 0^+$ is implicitly understood. To simplify the writing,
we set $ \hbar = m = 1 $ in the following formulas. In practice, this
amounts to measuring energies in units of $\fm^{-2}$ instead of MeV
with the conversion factor $ 1~\fm^{-2} = (\hbar^{2}/m)\;\fm^{-2} = 41.44\;\MeV$. In the present case that $\xi_k$ is
quadratic in $k$, the angular integral in \Eq{eq_lip_sch_4} can be
easily written with the angle-averaged two-particle Green's function
\begin{equation}\label{def_G_bar}
  \bar{G}_{0}^{(2)} (p, k, \omega)
  = \frac{1}{2}\int_{-1}^{1} \dd\cos \theta \, G_{0}^{(2)} (\pp, \kk, \omega)
  = \dfrac{\bar{Q}(k, p)}{\omega + 2 \mu - \tfrac{k^{2}}{4} - p^{2} + \ii \eps}
\end{equation}
where $\theta$ is the angle between $\kk$ and $\pp$ and the
angle-averaged Pauli blocking factor is given by
\begin{equation}\label{def-Qbar}
  \bar{Q}(k, p) = \frac{1}{2}\int_{-1}^{1} \dd\cos \theta\,
  \big[1 - \ff(\xi_{\kk/2 + \pp}) - \ff(\xi_{\kk/2 - \pp})\big]
 = \dfrac{2 T}{k p}
  \ln \left( \dfrac{1 + \e^{[(k/2+p)^2 - 2\mu]/(2T)}}
      {1 + \e^{[(k/2-p)^2 - 2\mu]/(2T)}}\right) - 1\,.
\end{equation}
In this way, \Eq{eq_lip_sch_4} reduces to a one-dimensional integral equation
\begin{equation}\label{eq_lip-schw}
\Gamma_{l=0}(q, q', k, \omega) = v_{l=0}(q, q') + \dfrac{2}{\pi}
\int_{0}^{\infty} \dd p \, p^{2}\, v_{l=0}(q, p)
\dfrac{\bar{Q}(k, p)}{p_{0}^{2}-p^{2}+\ii\eps} \Gamma_{l=0}(p,q',k,\omega),
\end{equation}
where
\begin{equation}\label{def_p0}
p_{0} = \sqrt{\omega + 2 \mu - k^{2} / 4}
\end{equation}
denotes the on-shell momentum in the center-of-mass frame. The
subscripts $l=0$ in \Eq{eq_lip-schw} indicate that we use
conventions that are common if one works in a partial wave basis,
i.e., $v = 4\pi v_{l=0}$ and $\Gamma = 4\pi\Gamma_{l=0}$.

%%%%%%%%%%%%%%%%%%%%%%%%%%%%%%%%%%%%%%%%%%%%%%%%%%%%%%%%%%%%%%%%%%%%%%%%%%
\subsection{Numerical solution for a non-separable interaction}
\label{subsec:nonseparable}
%%%%%%%%%%%%%%%%%%%%%%%%%%%%%%%%%%%%%%%%%%%%%%%%%%%%%%%%%%%%%%%%%%%%%%%%%%
In order to solve numerically \Eq{eq_lip-schw}, we will use
Weinberg's eigenvalues \cite{Weinberg1963,Ramanan2013}. If the
integral equation is schematically written as $\Gamma = V+VG\Gamma$,
the idea is to diagonalize the operator $ VG $, i.e., to find the
eigenvectors $u$ and eigenvalues $\eta$ such that $ V G u = \eta
u$. Explicitly, one has to solve for each $k$ and $\omega$
\begin{equation}\label{diag_op}
  \dfrac{2}{\pi} \int_{0}^{p_{\max}} \dd p \, p^{2} \,
  v_{l=0}(q, p) \dfrac{\bar{Q}(k, p)}{p_0^2-p^2 + \ii \eps} u_n(p, k, \omega)
  = \eta_n(k, \omega) u_n(q, k, \omega)\,.
\end{equation}
Here, we have assumed that the matrix elements $v(q,q')$ fall off fast
enough so that in practice the integral can be cut off at some
momentum $p_{\max}$.

To solve \Eq{diag_op}, we discretize the interval from $0$ to
$p_{\max}$ into a grid of $ n_q $ points $q_i$. Assuming that the
eigenfunctions $u_n$ can be interpolated between these points with a
cubic spline, we can write them as
\begin{equation}\label{def_u_chap}
u_n(q, k, \omega) = \sum_{i} c_{ni}(k, \omega) \varphi_{i}(q)\,,
\end{equation}
where $\varphi_{i}(q)$ is the piecewise cubic basis function for the
B-spline representation \cite{ToernigSpelucci} of $u_n$, i.e., it is a
``hat function'' having its maximum at $q_i$ and vanishing outside the
interval $(q_{i-2},q_{i+2})$. The advantage of this method is that,
although the kernel of the integral equation may be strongly peaked
near the Fermi surface (especially at low temperature), it is not
necessary to use a very fine momentum grid $q_i$, because the shape of
$u(q,k,\omega)$ follows the smooth $q$ dependence of $v_{l=0}(q,p)$.

By injecting the expression \eqref{def_u_chap} into \Eq{diag_op} and
taking the values of $q$ on the grid points $q_{i}$, we get
\begin{equation}\label{eq_int_wein}
  \sum_{j} \dfrac{2}{\pi} \int_{q_{j-2}}^{q_{j+2}} \dd p
  \, p^{2} \, v(q_{i}, p)
  \dfrac{\bar{Q}(k, p)}{p_0^2 - p^2 + \ii \eps} \, \varphi_{j}(p) \,
  c_{nj}(k, \omega)
= \eta_n(k, \omega) \sum_{j} \varphi_{j}(q_{i})\, c_{nj}(k, \omega)\,.
\end{equation}
Introducing the matrices
\begin{equation}
  M_{ij}(k, \omega)
  = \dfrac{2}{\pi} \displaystyle \int_{q_{j-2}}^{q_{j+2}} \dd p
  \, p^{2} \, v(q_{i}, p)
    \dfrac{\bar{Q}(k, p)}{p_{0}^{2} - p^{2} + \ii \eps} \varphi_{j}(p)
    \qquad\mbox{and}\qquad
  A_{ij} = \varphi_{j}(q_{i})\,,\label{defMA}
\end{equation}
we can write \Eq{eq_int_wein} in matrix notation as $M c_n = \eta_n A
c_n$ or finally
\begin{equation}
A^{-1}\, M(k,\omega)\, c_n(k,\omega) = \eta_n(k,\omega)\, c_n(k,\omega)\,,
\end{equation}
which is just an ordinary matrix diagonalization problem. In practice,
the imaginary part of the matrix $M$ can be calculated analytically,
\begin{equation}
  \Im M_{ij}(k,\omega) = -2\,v_{l=0}(q_i,p_0)\, \bar{Q}(k,p_0)\,\varphi_j(p_0)\,,
\end{equation}
and the real part is obtained as a principal-value integral.

Let us now express also the vertex function $ \Gamma $ in the basis of
the hat functions:
\begin{equation}\label{def_gamma_chap}
\Gamma_{l=0}(q, q', k, \omega) = \sum_{m, n} C_{mn}(k, \omega)\,
\varphi_{m}(q)\, \varphi_{n}(q')\,.
\end{equation}
Inserting this into \Eq{eq_lip-schw} and placing the points
$ q $ and $ q'$ on the grid points $ q_{i} $ and $q_{j}$, we get
\begin{equation}\label{eq_sum_mat}
  \sum_{m, n} C_{mn} \, \varphi_{m}(q_{i}) \, \varphi_{n}(q_{j})
  = v(q_{i}, q_{j})
  + \sum_{m, n} \dfrac{2}{\pi} \int_{q_{m-2}}^{q_{m+2}} \dd p
  \, p^{2}\, v(q_{i}, p)
  \dfrac{\bar{Q}(k, p)}{p_{0}^{2} - p^{2} + \ii \eps}
  \varphi_{m}(p) \, C_{mn}\, \varphi_{n}(q_{j})\,.
\end{equation}
We recognize the matrices $ M $ and $ A $ defined in
  \Eq{defMA}. We also define the matrix $ V_{ij} = v_{l=0}(q_{i},
q_{j})$, so that \Eq{eq_sum_mat} can be written in matrix notation as
$A C A^{\transpose} = V + M C A^{\transpose}$ and therefore
\begin{equation}
  C = (I - A^{-1} M)^{-1} \tilde{V}\,,
\end{equation}
with $ I $ the identity matrix and $ \tilde{V} = A^{-1} V
(A^{\transpose})^{-1}$ the coefficients one needs to express
$v_{l=0}(q,q')$ in terms of hat functions analogously to
\Eq{def_gamma_chap}.

Now we make use of Weinberg's eigenvalues $\eta_n$ and eigenvectors
$c_n$ which we determined previously. This allows us to write the
matrix $ A^{-1} M $ in the form $ A^{-1} M = P D P^{-1} $, with $D =
\text{diag} (\eta_{1}, \eta_{2}, \dots, \eta_{n_q})$ and $P = (c_{1},
c_{2}, \dots, c_{n_q})$. For given values of $k$ and $\omega$, the
coefficients $C_{ij}$ of the vertex function are given by
\begin{equation}\label{coef_fn_vertex}
C = P\, (I - D)^{-1} \, P^{-1} \, \tilde{V}\,.
\end{equation}

The vertex function allows us to determine the critical temperature $
T_{c} $. The Thouless criterion~\cite{Thouless1960} tells us that at
the critical temperature, a pole appears in the T matrix. This amounts
to solving the equation:
\begin{equation}
  \Gamma^{-1} (q, q', k=0, \omega = 0; T = T_{c} ) = 0\,.
\end{equation}
However, we can also determine $T_c$ directly from the Weinberg
eigenvalues as explained in~\cite{Ramanan2013}. When we look at the
coefficients given by \Eq{coef_fn_vertex}, we notice that the
pole in the vertex function appears if at least one of the eigenvalues
is equal to one. More explicitly, the critical temperature is reached
when
\begin{equation}
  \max \eta_{n} (k = 0, \omega = 0; T = T_{c}) = 1\,.
  \label{eq:Tc-Weinberg}
\end{equation}
One can easily see that this is equivalent to writing the BCS gap
equation in the limit of a vanishing gap, when the momentum dependent
gap becomes proportional to the eigenvector $\Delta_q \propto u_n(q,
0, 0)$.

%%%%%%%%%%%%%%%%%%%%%%%%%%%%%%%%%%%%%%%%%%%%%%%%%%%%%%%%%%%%%%%%%%%%%%%%%%
\subsection{Vertex function with a separable interaction}
%%%%%%%%%%%%%%%%%%%%%%%%%%%%%%%%%%%%%%%%%%%%%%%%%%%%%%%%%%%%%%%%%%%%%%%%%%
Solving~\Eq{eq_lip-schw} is much simpler if we consider the case of
a separable interaction, i.e.,
\begin{equation}\label{v_sep}
v(q, q') = g\, F(q)\, F(q')\,,
\end{equation}
where $ F $ is a form factor of the interaction and $ g $ the coupling
constant. (Separable interactions are also very helpful in
solving the gap equation below $T_c$ \cite{Khodel1996}.) In this
case, the vertex function can be written as
\begin{equation}\label{Gamma-separable}
  \Gamma (q, q', k, \omega) = \dfrac{F(q) \, F(q')}{1/g - J(k, \omega)} \,,
\end{equation}
with
\begin{equation}\label{def-J}
  J(k, \omega) = \dfrac{2}{\pi} \int_{0}^{\infty} \dd p\, p^{2}\, [F(p)]^2
  \dfrac{\bar{Q}(k, p)}{p_{0}^{2} - p^{2} + \ii \eps}\,,
\end{equation}
where $ p_{0} $ is defined in~\Eq{def_p0}.

The $ V_{\lowk} $ interaction with a cutoff of $\Lambda = 2\,
\fmi$ is reasonably well reproduced with a Gaussian form factor
\begin{equation}
F(q) = \e^{-q^{2}/q_{0}^{2}}\,.
\end{equation}
The two parameters to be adjusted are then the coupling constant $ g $
and the width of the Gaussian $ q_{0} $. One way to obtain these
parameters, given in Table~\ref{tab_fit},
%%%%%%%%%%%%%%%%%%%%%%%%%%%%%%%%%%%%%%%%%%%%%%%%%%%%%%%%%%%%%%%%%%%%%%%%%%%%%%%
\begin{table}
\begin{center}
\renewcommand{\arraystretch}{1.3} 
\setlength{\tabcolsep}{0.48cm}
\begin{tabular}{c|c|c}
\hline 
\hline 
 & $ g_{l=0} \; (\fm) $ & $ q_{0} \; (\fmi) $  \\
\hline
$nn$ interaction& $ -1.644 $& $ 1.367 $ \\
unitary limit &  $ -1.834 $ & $ 1.367 $  \\
\hline 
\hline 
\end{tabular} 
\caption{Parametrizations of the separable interaction. First line:
  fit to the BCS critical temperature computed with $ V_{\lowk} $ with
  cut-off $ \Lambda = 2 \, \fmi $ [$g_{l=0}$ given here is related to
    $ g = -856\;\MeV\, \fm^{3}$ in~\cite{Martin2014} by $g_{l=0} = g
    m/(4\pi \hbar^2)$]. Second line: unitary limit ($a\to\infty$).
\label{tab_fit}}
\end{center}
\end{table}
%%%%%%%%%%%%%%%%%%%%%%%%%%%%%%%%%%%%%%%%%%%%%%%%%%%%%%%%%%%%%%%%%%%%%%%%%%%%%%%
is to fit the critical temperature curve obtained with the $ V_{\lowk}
$ interaction in the BCS approximation \cite{Martin2014}, as shown in
\Fig{Tc-mu_cut-off}. In view of the quality of the fit and the gain in
simplicity of the calculations, we will perform some of our
calculations with this separable interaction instead of the
$V_{\lowk}$ interaction.

With the parameters given in Table~\ref{tab_fit}, one finds a
scattering length $ a = -15.9 \;\fm$ which is a bit smaller than the
real $nn$ scattering length \cite{Gonzales1999,Babenko2013} but still
very large compared to the range of the $nn$ interaction and therefore
close to the unitary limit $a\to\infty$. Although in nature, neutron
matter is not in the unitary limit, it is interesting to consider it
as a test case for the theory. This can be done by slightly
readjusting the coupling constant, see Table~\ref{tab_fit}.

%%%%%%%%%%%%%%%%%%%%%%%%%%%%%%%%%%%%%%%%%%%%%%%%%%%%%%%%%%%%%%%%%%%%%%%%%%%%%%%%
\subsection{Self-energy}
%%%%%%%%%%%%%%%%%%%%%%%%%%%%%%%%%%%%%%%%%%%%%%%%%%%%%%%%%%%%%%%%%%%%%%%%%%%%%%%%
The Feynman diagram for the self-energy in ladder approximation is
represented in \Fig{fig:feynman}(b). The corresponding expression
reads~\cite{Ramanan2013}
\begin{equation}\label{def_sigma_n}
  \Sigma(k, \ii \omega_{n}) = \int \dfrac{\dd^{3} \pp}{(2 \pi)^{3}}\,
  \dfrac{1}{\beta} \sum_{m} G_{0}(p, \ii \omega_{m})
  \Gamma\big(\tfrac{|\kk - \pp|}{2}, |\kk + \pp|,
    \ii \omega_{n} + \ii \omega_{m}\big)\,.
\end{equation}
Since here in- and outgoing momenta in $\Gamma$ are equal, we use the
notation $\Gamma(q,k,\omega)$ for $\Gamma(q,q,k,\omega)$. The
self-energy can be split into two contributions $\Sigma(k, \omega) =
\Sigma_{\HF} (k) + \Sigma_{\lad} (k, \omega)$, where $\Sigma_{\HF}$ is
the energy-independent Hartree-Fock (HF) term
\begin{equation}\label{Sigma_HF}
  \Sigma_{\HF} (k) = \int \dfrac{\dd^{3} \pp}{(2 \pi)^{3}} \,
  v\big( \tfrac{|\kk - \pp|}{2}, \tfrac{|\kk - \pp|}{2} \big)\, \ff(\xi_{p})\,,
\end{equation}
generated by the first-order (Born) term in $\Gamma$, and
$\Sigma_{\lad}$ is the remaining energy dependent term. After
analytic continuation to real energies, the imaginary part of the
retarded self-energy can be written as
\begin{equation}\label{ImSigma}
  \Im \Sigma(k, \omega) =
  - \int \dfrac{\dd^{3} \pp}{(2 \pi)^{3}}\, [\ff(\xi_{p}) + \fb(\omega+\xi_{p})]
\Im \Gamma(\tfrac{|\kk - \pp|}{2}, |\kk + \pp|, \omega + \xi_{p})\,,
\end{equation}
where $\fb(\xi) = 1/(\e^{\beta\xi}-1)$ denotes the Bose function. The
real part can then be computed as
\begin{equation}
  \label{Sigma_spectral}
  \Re \Sigma (k, \omega) = \Sigma_{\HF}(k) -
  \principalvalue \int \dfrac{\dd \omega'}{\pi} \,
  \dfrac{\Im \Sigma(k, \omega')}{\omega - \omega'}\,,
\end{equation}
where $\mathcal{P}$ denotes the principal value.

%%%%%%%%%%%%%%%%%%%%%%%%%%%%%%%%%%%%%%%%%%%%%%%%%%%%%%%%%%%%%%%%%%%%%%%%%%%%%%%
\subsection{Occupation numbers}
\label{subsec:occ-numbers-formalism}
%%%%%%%%%%%%%%%%%%%%%%%%%%%%%%%%%%%%%%%%%%%%%%%%%%%%%%%%%%%%%%%%%%%%%%%%%%%%%%%
The BCS-BEC crossover (strong-coupling) regime is characterized by the
existence of pairing correlations even above the superfluid critical
temperature. These modify the occupation numbers which are therefore
no longer given by those of an ideal Fermi gas, $n_0(k) = \ff(\xi_k)$.
The presence of non-condensed pairs can considerably modify the
relationship between the chemical potential $\mu$ and the density and
thus the density dependence of the critical temperature $T_c$.

The first way to include this effect is the Nozi\`{e}res--Schmitt-Rink
(NSR) approach \cite{Nozieres1985,SadeMelo1993}. Although originally
formulated differently, it amounts to computing the density from a
Green's function which is dressed with only one self-energy insertion
\cite{CalvaneseStrinati2018}, i.e.,
\begin{equation}\label{G_NSR}
  G(k, \ii\omega_{n}) = G_{0}(k, \ii\omega_{n}) + [G_{0}(k, \ii\omega_{n})]^2 \,
  \Sigma(k, \ii \omega_{n})\,.
\end{equation}
Inserting for $\Sigma$ the spectral representation (right-hand side of
\Eq{Sigma_spectral} without the principal value and with $\omega$
replaced by $\ii\omega_n$), and performing the summations over
$\omega_n$, one obtains the NSR occupation numbers $ n_{\NSR}(k) =
n_0(k) + n_{\NSR}^{c} (k)$, with 
\begin{equation}\label{n_NSR}
  n_{\NSR}^{c}(k) = \ff '(\xi_{k}) \, \Sigma_{\HF} (k)
  + \int \dfrac{\dd \omega}{\pi} \, \Im \Sigma (k, \omega)
  \dfrac{\ff(\xi_{k}) - \ff(\omega) - (\xi_{k} - \omega)\ff ' (\xi_{k})}
        {(\xi_{k} - \omega)^{2}} \,.
\end{equation}
By $\ff'(\xi)$ we denote the derivative $d\ff(\xi)/d\xi$.

However, the occupation numbers obtained within the NSR approach are
not satisfactory (see for example \Fig{fig_nb_occupa_mu} (b)
below). The real part of the self-energy produces a shift of the
quasiparticle energy which the NSR theory improperly counts as part
of the correlation correction. For example, if we consider a constant
self-energy (of type $\Sigma_{\HF}$), \Eq{n_NSR} (of which only the
first term remains) is only a perturbative expansion of the occupation
numbers $n(k) = \ff(\xi_{k}^{\star})$ of an uncorrelated gas of
quasiparticles of energy $\xi_{k}^{\star } = \xi_{k} +
\Sigma_{\HF}(k)$.

This problem has been solved for the case of symmetric nuclear matter
in \Refs{Schmidt1990,Stein1995,Jin2010} following an approach
presented initially for an electronic system in~\cite{Zimmermann1985}.
In simple words, we must eliminate the shift of the quasiparticle
energy from the calculation of the correlations. Within the formalism
of self-consistent Green's functions~\cite{Haussmann2007}, we should
compute the T matrix $\Gamma$ and the self-energy $\Sigma$ with the
dressed Green's function $G = 1 / (\omega - \xi_{k} - \Sigma)$.  The
quasiparticle approximation consists in approximating in $\Gamma$ and
$\Sigma$ this Green's function by $G_{0}^{\star} = 1 / (\omega -
\xi_{k}^{\star})$, with
\begin{equation}\label{xi_etoile_auto}
\xi_{k}^{\star} = \xi_{k} + \Re \Sigma (k, \xi_{k}^{\star})\,,
\end{equation}
where $ \Sigma $ should be computed self-consistently with
$G_{0}^{\star}$. Thus, the dressed Green's function can be written as
\begin{equation}\label{g_habille_nb}
G(k, \omega) = \dfrac{1}{\omega - \xi_{k} - \Sigma(k, \omega)} =
\dfrac{1}{\omega - \xi_{k}^{\star} - [\Sigma(k, \omega) - \Re \Sigma
    (k, \xi_{k}^{\star})]} \; \, ,
\end{equation}
which shows that one has to subtract $\Re \Sigma(k,\xi_{k}^{\star})$
from the self-energy. In practice, as our self-energy contains only
the $s$-wave part of the total interaction, we do not have the means
to calculate $ \xi_{k}^{\star} $. We could for example use an
energy-density functional of Skyrme or Gogny type to calculate the
effective mass and the shift of the chemical potential. In neutron
matter, the effective mass is close to the free
mass~\cite{Buraczynsk2019}. The rest can be absorbed in a shift in
chemical potential which does not impact our results because we do not
calculate thermodynamic quantities. For this reason, we directly use
$\xi_{k}^{\star} = \xi_{k}$.

A first approximation presented in~\cite{Ramanan2013} consists in
subtracting only the HF part from the self-energy. The HF self-energy
is expected to be the dominant term, and this avoids the computation
of the complete self-energy. Furthermore, analogous to the
approximation \eqref{G_NSR} used in NSR theory, the dressed Green's
function \eqref{g_habille_nb} is again truncated to first order. Thus,
we have the following expression for the dressed Green's function:
\begin{equation}
G(k, \ii\omega_{n}) = G_{0}(k, \ii\omega_{n}) + [G_{0}(k, \ii\omega_{n})]^2 \,
[\Sigma(k, \ii \omega_{n}) - \Sigma_{\HF}(k)]\,.
\end{equation}
Analogously to the NSR case, we get now $ n_{\sHF}(k) = n_{0} (k) +
n_{\sHF}^{c} (k)$, with
\begin{equation}\label{n_s-HF}
n_{\sHF}^{c}(k) = \int\dfrac{\dd \omega}{\pi}\, \Im \Sigma (k, \omega)\,
\dfrac{\ff(\xi_{k}) - \ff(\omega) - (\xi_{k} - \omega)\ff'(\xi_{k})}
      {(\xi_{k} - \omega)^{2}}\,.
\end{equation}

However, in contrast to \Reference{Ramanan2013}, we do compute the
self-energy and therefore nothing prevents us from subtracting the
total self-energy instead of only the HF part. In this case, we write
\begin{equation}
  G(k, \ii\omega_{n}) = G_{0}(k,\ii \omega_{n}) + [G_{0}(k, \ii \omega_{n})]^2 \,
  [\Sigma(k,  \ii \omega_{n}) - \Re \Sigma(k, \xi_{k})]\,,
\end{equation}
which leads to the occupation numbers $n_{\stot}(k) = n_{0}(k) +
n_{\stot}^{c}(k)$, with \cite{Jin2010}
\begin{equation}\label{n_s-tot}
  n_{\stot}^{c}(k) = \principalvalue \int
  \dfrac{\dd \omega}{\pi} \, \Im \Sigma (k, \omega) \,
  \dfrac{\ff(\xi_{k}) - \ff(\omega)}{(\xi_{k} - \omega)^{2}}\,.
\end{equation}

In the three methods above, the correlated occupation numbers are
calculated from the first-order truncated Dyson equation. But we can
also consider the complete Dyson equation. In this case, we compute
the spectral function as the imaginary part of the Green's function
\begin{equation}\label{def_fun_spect_T_fini}
  A(k, \omega) = - \Im \dfrac{1}{\omega - \xi_{k} - [\Sigma(k, \omega) -
      \Re \Sigma(k, \xi_{k})]}\,,
\end{equation}
which allows us to calculate the occupation numbers $ n_{\D} $ with
the complete Dyson equation
\begin{equation}\label{n_D}
n_{\D}(k) = \int \dfrac{\dd\omega}{\pi} A(k, \omega) \, \ff(\omega) \, .
\end{equation}

%%%%%%%%%%%%%%%%%%%%%%%%%%%%%%%%%%%%%%%%%%%%%%%%%%%%%%%%%%%%%%%%%%%%%%%%%%%%%%%
\section{Numerical results}
\label{sec:results}
%%%%%%%%%%%%%%%%%%%%%%%%%%%%%%%%%%%%%%%%%%%%%%%%%%%%%%%%%%%%%%%%%%%%%%%%%%%%%%%

%%%%%%%%%%%%%%%%%%%%%%%%%%%%%%%%%%%%%%%%%%%%%%%%%%%%%%%%%%%%%%%%%%%%%%%%%%%%%%%
\subsection{Critical temperature as a function of the chemical potential}
\label{subsec:Tc-mu}
%%%%%%%%%%%%%%%%%%%%%%%%%%%%%%%%%%%%%%%%%%%%%%%%%%%%%%%%%%%%%%%%%%%%%%%%%%%%%%%
As pointed out in the end of \Sec{subsec:nonseparable}, the critical
temperature $T_c$ determined via \Eq{eq:Tc-Weinberg} as a function of
the chemical potential $\mu$ is the same as in the BCS
approximation. The difference appears only when one computes $T_c$ as
a function of the density $n$, since the correlations change the
relationship between $\mu$ and $n$.

\begin{figure}
\begin{center}
\includegraphics[width=7.5cm]{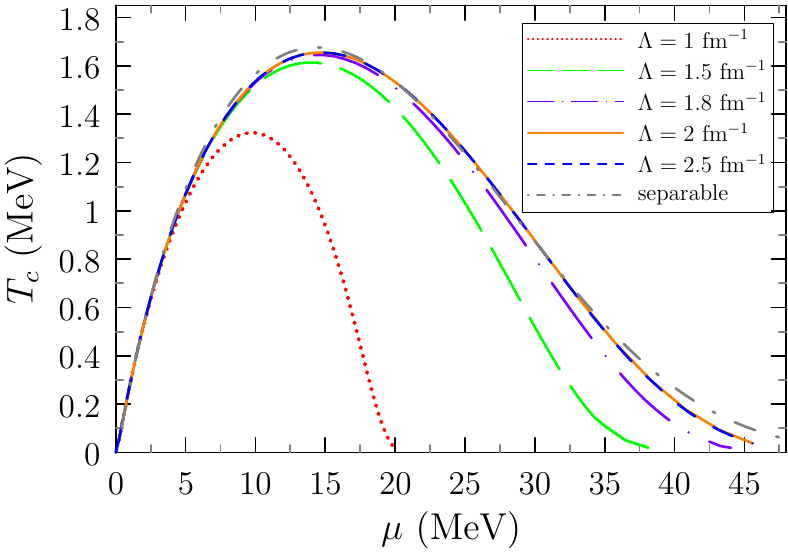}
\caption{BCS critical temperature $ T_{c} $ as a function of the
  chemical potential $\mu$ for different cutoff values
  $\Lambda$. The red, green, purple, orange and blue curves correspond
  to $ \Lambda = 1$, $1.5$, $1.8$, $2$, and $2.5\;\fmi$,
  respectively. The curves obtained for smaller cutoffs are only valid
  in the range where they agree with the $\Lambda = 2$ and
  $2.5\;\fmi$ results. The gray curve represents the critical
  temperature obtained with the separable potential \eqref{v_sep}.}
\label{Tc-mu_cut-off}
\end{center}
\end{figure}

In \Fig{Tc-mu_cut-off},
%%%%%%%%%%%%%%%%%%%%%%%%%%%%%%%%%%%%%%%%%%%%%%%%%%%%%%%%%%%%%%%%%%%%%%%%%%%%%%%
%%%%%%%%%%%%%%%%%%%%%%%%%%%%%%%%%%%%%%%%%%%%%%%%%%%%%%%%%%%%%%%%%%%%%%%%%%%%%%%
we display $T_c$ as a function of $\mu$, computed with different
$V_{\lowk}$ interactions. Matrix elements of the $V_{\lowk}$
interaction are available for arbitrary values of the cutoff
$\Lambda$. In our calculations, we use matrix elements of
\Reference{Ramanan2018}. For each cutoff, the $V_{\lowk}$ interaction is
constructed such that it reproduces the low-energy $nn$ scattering
data, but of course only at momenta below that cutoff. Thus the lower
cutoffs make only sense for small values of the chemical potential
where the cutoff dependence does not yet set in. For $\Lambda \geq
2\;\fmi$, the $T_c$ vs. $\mu$ curve remains unchanged over the
whole range of chemical potentials and shows the usual behaviour. For
the smaller cutoffs, however, we see that the chemical potential must
be limited to $\mu < 5\;\MeV$ for $\Lambda = 1\;\fmi$, $\mu <
10\;\MeV$ for $\Lambda = 1.5\;\fmi$, and so on, which can be
summarized as $\mu[\MeV] \lesssim 4 \, (\Lambda[\fmi])^{2}$. We
also display the result obtained with the separable interaction. For
$\mu\lesssim 35\;\MeV$, it reproduces well the $V_{\lowk}$ results
obtained with $\Lambda=2\;\fmi$.

%%%%%%%%%%%%%%%%%%%%%%%%%%%%%%%%%%%%%%%%%%%%%%%%%%%%%%%%%%%%%%%%%%%%%%%%%%%%%%%%
\subsection{Spectral function and pseudogap}
\begin{figure}
\begin{center}
\includegraphics[width=7.5cm]{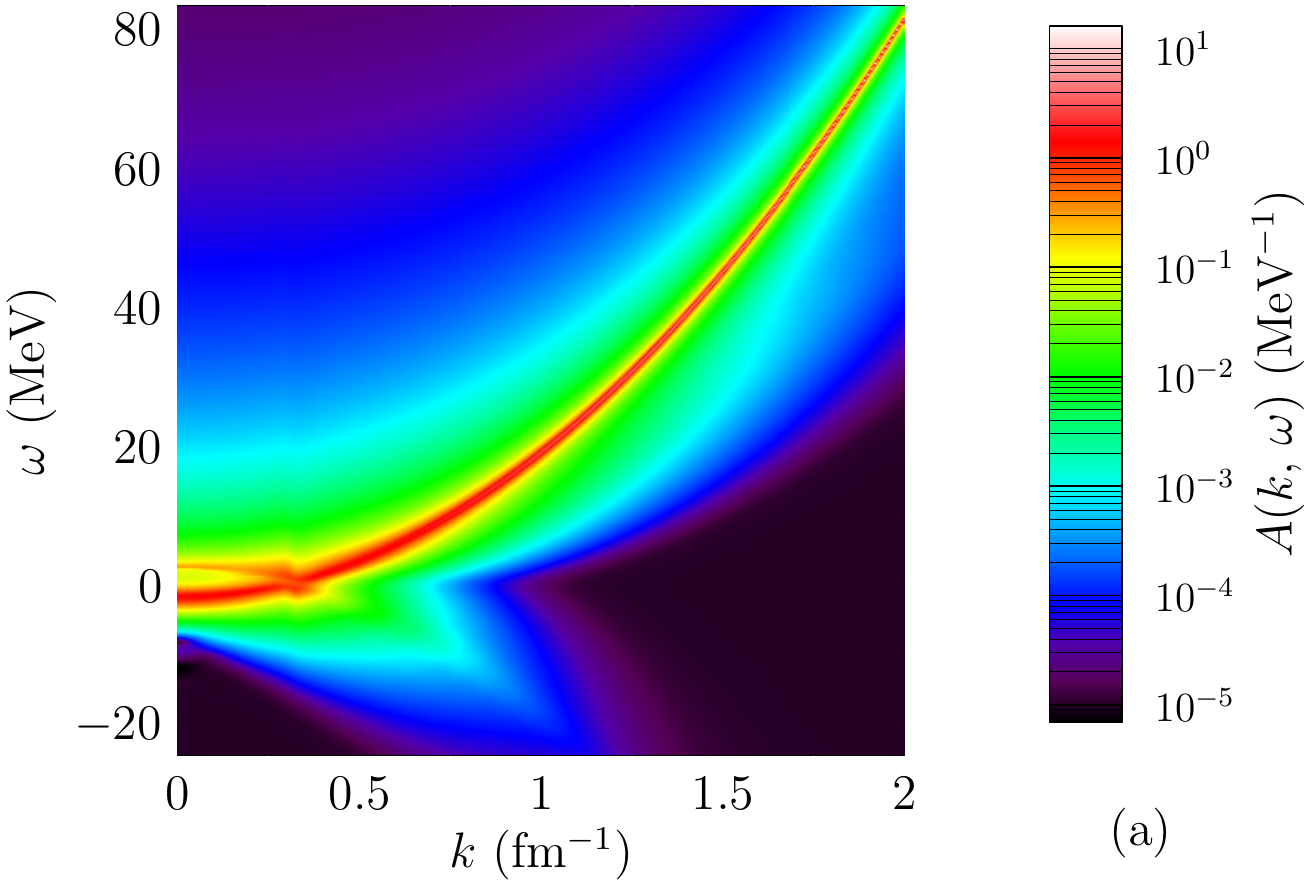} \hspace*{0.4cm}
\includegraphics[width=7.5cm]{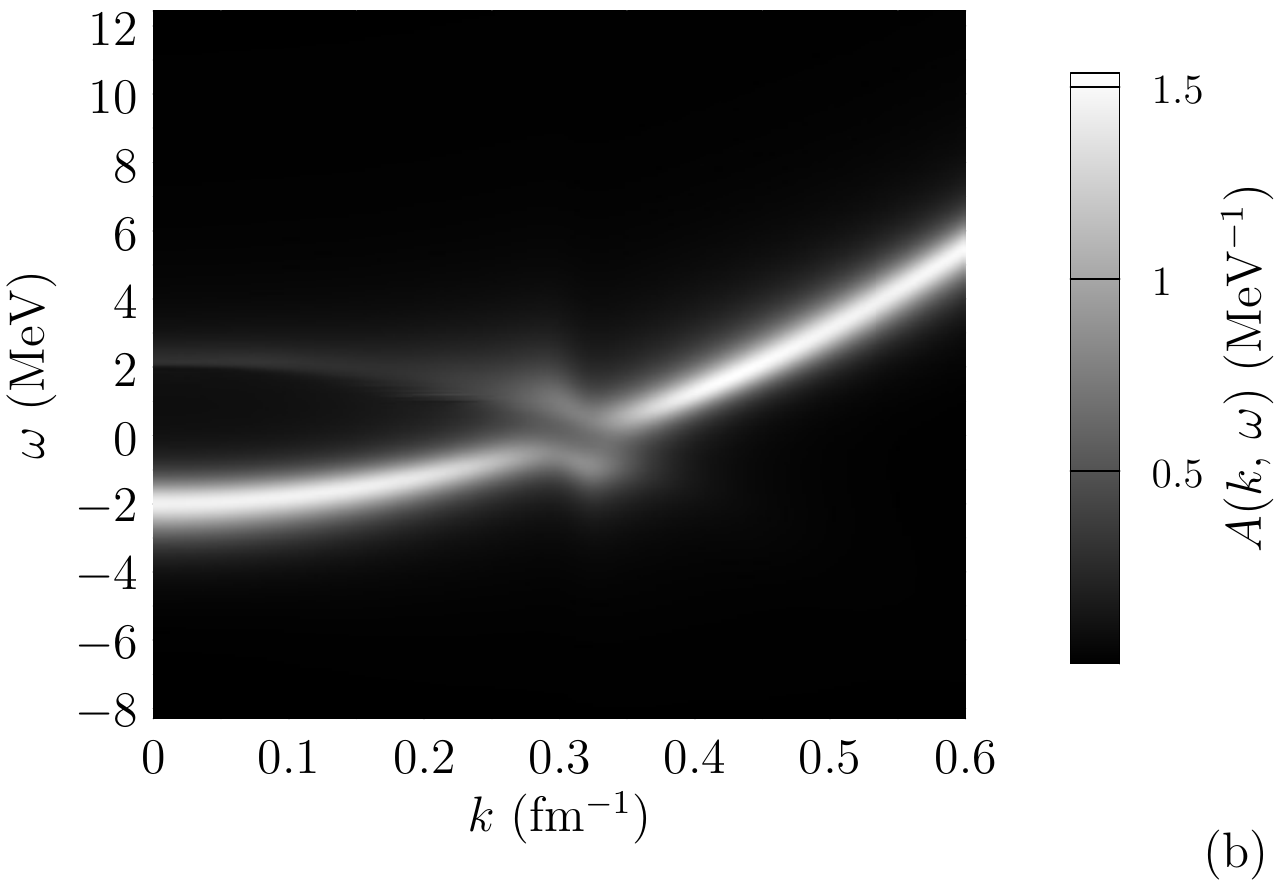}
\caption{Two-dimensional map of the spectral function $A(k,\omega)$
  for $ \mu = 2\;\MeV$ and $ T = 1.01 \, T_{c} (\mu) $ with the
  separable interaction. In figure (a), we can clearly see the
  position of the peak located at $ \omega = k^{2} / 2 -\mu $ and a
  double-peak structure for $ k \lesssim k_{\F} $. Figure (b) is a
  zoom on the region with the double peak.}
\label{graphe_fonction_spectrale_2d}
\end{center}
\end{figure}

%%%%%%%%%%%%%%%%%%%%%%%%%%%%%%%%%%%%%%%%%%%%%%%%%%%%%%%%%%%%%%%%%%%%%%%%%%%%%%%%
Below the critical temperature $T_{c}$, the pairing gap corresponds to
an energy interval around the Fermi energy ($\omega=0$) where there
are no states, because adding or removing a particle requires the
energy to break a pair. Above $T_{c}$, such a gap does not exist, but
in some cases one observes an energy region of reduced density of
states, which is called the pseudogap. The spectral function
$A(k,\omega)$, \Eq{def_fun_spect_T_fini}, can give give us information
on this pseudogap. Figure~\ref{graphe_fonction_spectrale_2d}
%%%%%%%%%%%%%%%%%%%%%%%%%%%%%%%%%%%%%%%%%%%%%%%%%%%%%%%%%%%%%%%%%%%%%%%%%%%%%%%%
%%%%%%%%%%%%%%%%%%%%%%%%%%%%%%%%%%%%%%%%%%%%%%%%%%%%%%%%%%%%%%%%%%%%%%%%%%%%%%%%
displays the spectral function computed for $\mu = 2\;\MeV$ at $ T =
0.53 \;\MeV$ which is slightly above $T_c$. We can clearly see a
double peak for momenta below and around $k_{\F}$, as well as the
significant level repulsion between the two peaks near $k =
k_{\F}$. Although we are in the normal phase, this behavior reminds
the quasiparticle spectrum in BCS theory in the superfluid phase below
$T_c$, except that the peaks have a width and the spectral function
does not vanish between the peaks.

\begin{figure}
\begin{center}
\includegraphics[width=7.5cm]{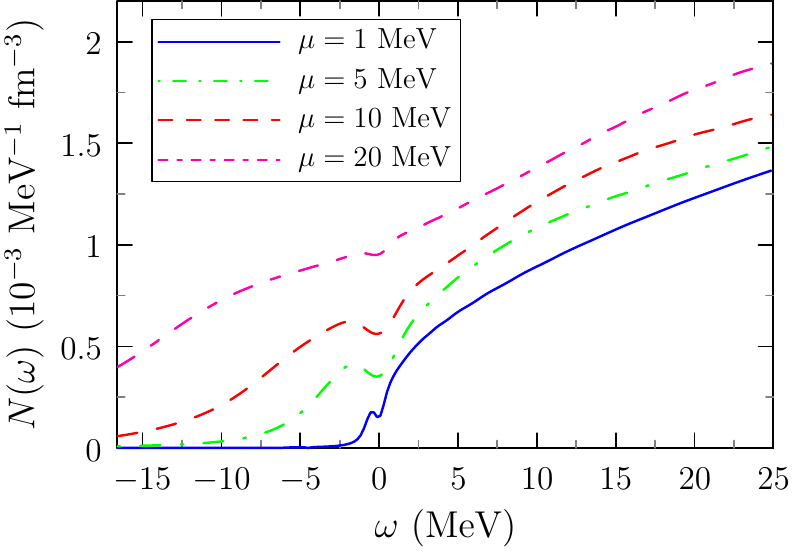}
\caption{Level densities for several values of the chemical potential
  $ \mu = 1\;\MeV$ (blue line), $5\;\MeV$ (green dash-dot line),
  $10\;\MeV$ (red dashed line) and $20\;\MeV$ (double-dashed purple
  line) slightly above the respective critical temperatures $
  T = 1.01 \,T_{c}(\mu)$ which are $ 0.27, 1.09, 1.57$, and $1.53\; \MeV $.}
\label{graphe_pseudo_gap}
\end{center}
\end{figure}

To quantify the presence of the pseudogap, we define the level density
\begin{equation}
N(\omega) = \dfrac{1}{2 \pi^3} \int_{0}^{\infty} \dd k\, k^2\, A(k,\omega)
\end{equation}
Figure~\ref{graphe_pseudo_gap}
%%%%%%%%%%%%%%%%%%%%%%%%%%%%%%%%%%%%%%%%%%%%%%%%%%%%%%%%%%%%%%%%%%%%%%%%%%%%%%%%
%%%%%%%%%%%%%%%%%%%%%%%%%%%%%%%%%%%%%%%%%%%%%%%%%%%%%%%%%%%%%%%%%%%%%%%%%%%%%%%%
shows the level density calculated for different values of the
chemical potential at the corresponding critical temperatures. In all
cases, we see a dip in the level density around $\omega=0$. For
$\mu=20\;\MeV$, however, this reduction is very weak, because we are
no longer in the strong-coupling regime. For $\mu=1\;\MeV$ the dip
looks very small, but one has to remember that also the relevant
energy scale given by $\mu$ is much smaller.

While, to our knowledge, the pseudogap has not yet been discussed for
the case of pure neutron matter, we notice that it was theoretically
predicted for the case of dilute symmetric nuclear matter, which
undergoes a true crossover from a BEC of deuterons to a BCS phase of
$pn$ Cooper pairs \cite{Schnell1999}. In ultracold Fermi gases in the
BCS-BEC crossover, some hints for the backbending of the peak energy
near $k=k_{\F}$ at temperatures above $T_c$ were experimentally
observed \cite{Gaebler2010}. However, even in the unitary Fermi gas,
where pairing correlations are stronger than in neutron matter, it is
not clear whether the pseudogap phase exists. Theoretically, it tends
to be very visible in non self-consistent T-matrix approaches, such
as the present one, but less pronounced in self-consistent Green's
functions and quantum Monte-Carlo calculations \cite{Jensen2019}. A
pseudogap can have some effect on the spin susceptibility and the
specific heat \cite{Jensen2020}, but in the present case these effects
are probably too weak to be observed in neutron stars.

%%%%%%%%%%%%%%%%%%%%%%%%%%%%%%%%%%%%%%%%%%%%%%%%%%%%%%%%%%%%%%%%%%%%%%%%%%%%%%%%
\subsection{Occupation numbers}\label{subsec:occ-numbers-results}
\begin{figure}
\begin{center}
\includegraphics[width=7.5cm]{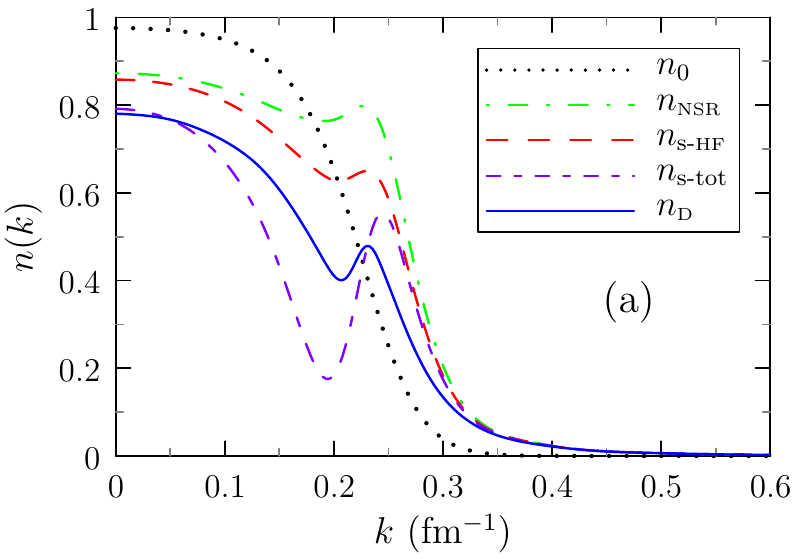} \hspace*{0.4cm}
\includegraphics[width=7.5cm]{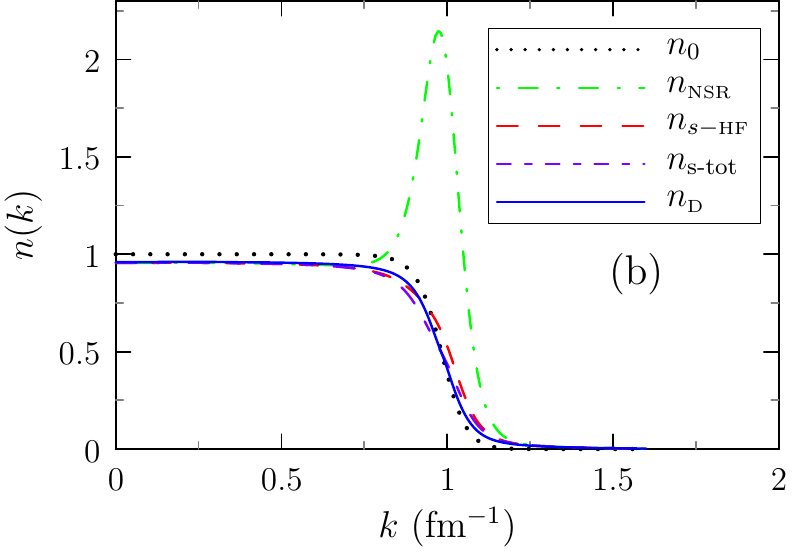}
\caption{Occupation numbers at $ T = 1.01 \, T_{c}(\mu) $ with (a) $
  \mu = 1\;\MeV$ and (b) $ \mu = 20\;\MeV$. The dotted black curve
  represents the free occupation numbers. The green dash-dot curve
  represents the NSR occupation numbers. The occupation numbers with
  HF subtraction (s-HF) and total subtraction (s-tot) are represented,
  respectively, by the red dashed curve and by the purple double
  dashed curve. Occupation numbers calculated with the full Dyson
  equation are represented by the blue line.}
\label{fig_nb_occupa_mu}
\end{center}
\end{figure}

%%%%%%%%%%%%%%%%%%%%%%%%%%%%%%%%%%%%%%%%%%%%%%%%%%%%%%%%%%%%%%%%%%%%%%%%%%%%%%%%
Figure~\ref{fig_nb_occupa_mu}
%%%%%%%%%%%%%%%%%%%%%%%%%%%%%%%%%%%%%%%%%%%%%%%%%%%%%%%%%%%%%%%%%%%%%%%%%%%%%%%%
%%%%%%%%%%%%%%%%%%%%%%%%%%%%%%%%%%%%%%%%%%%%%%%%%%%%%%%%%%%%%%%%%%%%%%%%%%%%%%%%
shows the shape of the occupation numbers, computed at the critical
temperature (in practice, slightly above $T_c$), for the different
approximations defined in \Sec{subsec:occ-numbers-formalism}. The left
panel (a) was computed with $\mu=1\;\MeV$, where the neutron matter is
in a strong-coupling regime since with $ T_c = 0.27 \; \MeV $ the
ratio $T_c/\mu$ is not small. We notice that the occupation numbers
including the correlations look all very different from the free ones
(dotted black line) but they depend strongly on the choice of the
approximation which is used (NSR, subtraction of the HF self-energy,
subtraction of the total self-energy, or full Dyson equation).

The right panel (b) was computed with $\mu = 20\;\MeV$. In this case,
the critical temperature of $ T_c = 1.52 \;
\MeV$ is close to its maximum, but the ratio $ T_c/\mu $ is much smaller
than for $\mu=1\;\MeV$, and we are therefore more in a weak-coupling
situation. In this case, one would expect that BCS theory is valid and
the occupation numbers above $T_c$ are close to the free ones. As we
see from the figure, this is indeed the case for all approximations
except NSR. The peak in the NSR occupation numbers is clearly
unphysical. When we look at the formula~\eqref{n_NSR}, we can see that
the derivative of the Fermi function will produce a peak near
$k_{\F}$. This has nothing to do with correlations but it simply
reflects the shift of the quasiparticle energies, treated to first
order in a Taylor expansion. This effect is absent in the other
approximations where the chemical potential has to be interpreted as
an effective one, $\mu^{\star}$, which already includes the shift, as
pointed out below \Eq{g_habille_nb}. Therefore, the NSR curve and the
other curves are not really comparable as they correspond to different
values of the real chemical potentials.

%%%%%%%%%%%%%%%%%%%%%%%%%%%%%%%%%%%%%%%%%%%%%%%%%%%%%%%%%%%%%%%%%%%%%%%%%%%%%%%%
\subsection{Correlated density}
\label{subsec:rhocorr-mu}
%%%%%%%%%%%%%%%%%%%%%%%%%%%%%%%%%%%%%%%%%%%%%%%%%%%%%%%%%%%%%%%%%%%%%%%%%%%%%%%%
The main objective of the T-matrix theory is to determine the
density dependence of the critical temperature. As mentioned in
\Sec{subsec:Tc-mu}, the relation between $T_c$ and $\mu$ is the same
as in BCS theory, and the effect of the correlations above $T_c$
enters only through the $\mu$ dependence of the density $\rho$. The
density is calculated directly from the occupation numbers as
\begin{equation}
\rho = 2 \int \dfrac{\dd^{3} k}{(2\pi)^{3}} \, n(k)\,,
\end{equation}
where the factor of two accounts for the spin degeneracy. As we have
seen in \Sec{subsec:occ-numbers-formalism}, for the approximations
that treat the correlations only perturbatively (NSR, HF subtraction
or total subtraction), the density is naturally separated into two
parts such that $ \rho = \rho_{0} + \rho_{c} $, with $ \rho_{0} $ the
free density and $\rho_{c}$ the correlated density. In the case of
the full Dyson equation, we directly obtain the total occupation
numbers and therefore the total density $\rho$. In this case, we
define the correlated density as $\rho_{c} = \rho-\rho_{0}$.

\begin{figure}
\begin{center}
\includegraphics[width=7.5cm]{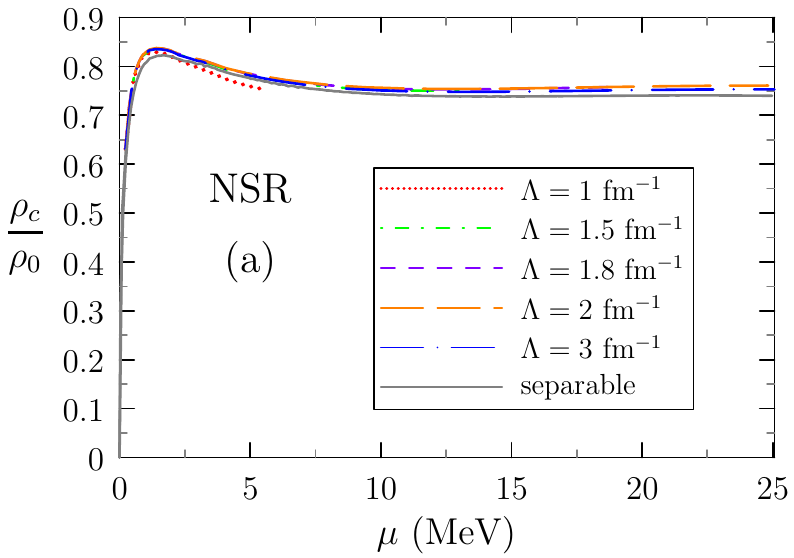} \hspace*{0.4cm}
\includegraphics[width=7.5cm]{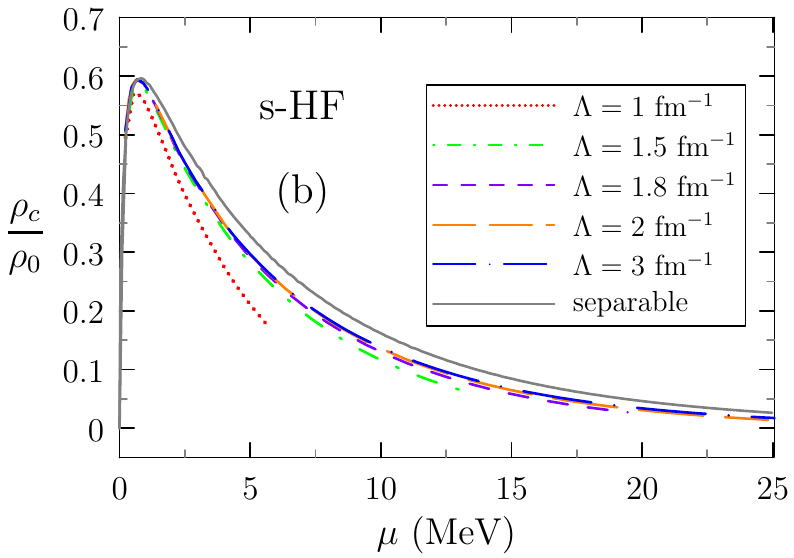} \\
\includegraphics[width=7.5cm]{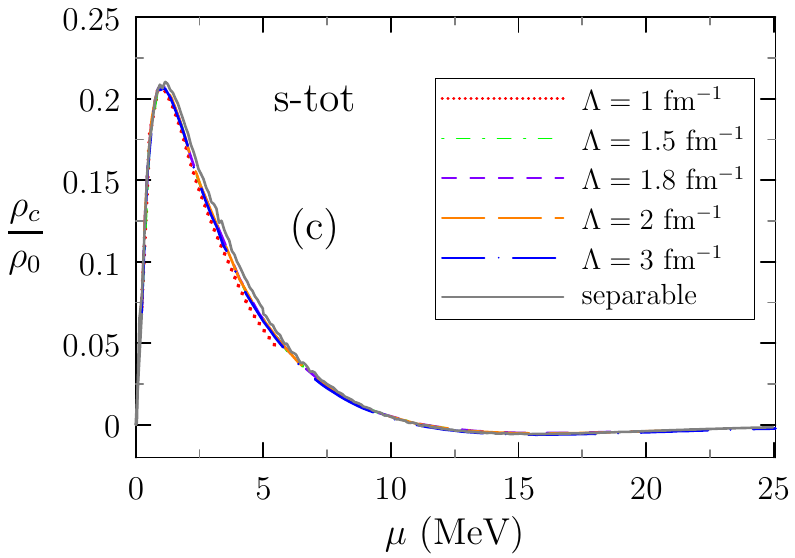} \hspace*{0.4cm}
\includegraphics[width=7.5cm]{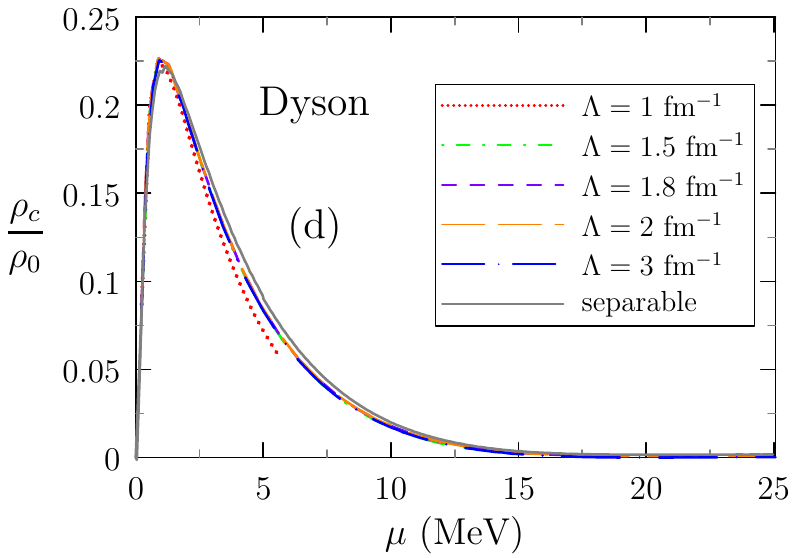}
\caption{Correlated densities at $ T = 1.01 \, T_{c}(\mu) $ as a
  function of the chemical potential for four methods to calculate the
  correlations: (a) NSR, (b) HF subtraction, (c) total self-energy
  subtraction, and (d) full Dyson equation. The calculations were done
  with $V_{\lowk}$ interactions corresponding to cutoffs $\Lambda = 1$
  (red), $1.5$ (green), $1.8$ (purple), $2$ (orange), $3$ (blue), and
  with the separable interaction (grey curves).}
\label{densite_cut-off}
\end{center}
\end{figure}

The four panels in \Fig{densite_cut-off} represent the ratios
$\rho_{c}/\rho_{0}$, computed at the critical temperature $T_c(\mu)$,
as a function of the chemical potential $\mu$, in the four different
approximations: (a) NSR, \Eq{n_NSR}; (b) HF subtraction, \Eq{n_s-HF};
(c) total subtraction, \Eq{n_s-tot}; and (d) full Dyson equation,
\Eq{n_D}.
%%%%%%%%%%%%%%%%%%%%%%%%%%%%%%%%%%%%%%%%%%%%%%%%%%%%%%%%%%%%%%%%%%%%%%%%%%%%%%%
%%%%%%%%%%%%%%%%%%%%%%%%%%%%%%%%%%%%%%%%%%%%%%%%%%%%%%%%%%%%%%%%%%%%%%%%%%%%%%%
The calculations were done using $V_{\lowk}$ interactions with
different cutoffs $\Lambda$ and with the separable
  interaction. Before we turn to a discussion of the cutoff
dependence, let us look at the results obtained with the largest
cutoffs $\Lambda = 2$ and $3\;\fm$ (orange and blue long
dashes), which lie practically on top of
each other and are very close to the results obtained with the
  separable interaction (gray solid lines).

We see that at very small chemical potential, the ratio
$\rho_c/\rho_0$ rises quickly, because the $nn$ scattering length is
so large that one quickly gets into the strong-coupling
regime. However, what happens at larger chemical potentials depends on
the approximation. Since the $nn$ interaction gets weaker at higher
momentum, one would expect that neutron matter returns into the
weak-coupling regime at high density, i.e., at high chemical
potentials, and in this case, the ratio $\rho_c/\rho_0$ should become
small. We see that this is true for all approximations except the NSR
one [\Fig{densite_cut-off}(a)]. This shows us once again the problem
that exists with the NSR theory, namely that it counts the effect of
the shift of the quasiparticle energy as ``correlations'' (or as
``fluctuations'' in the terminology of \Reference{Inotani2019}). Thus, in
the weak coupling limit at large values of the chemical potential, the
NSR ``correlated density'' remains very high whereas it should tend
towards zero.\footnote{We notice that our NSR results shown in
  \Fig{densite_cut-off}(a) agree with the ``uncorrected'' curve in
  Fig.~6 of \Reference{Ramanan2013}, computed directly from the Weinberg
  eigenvalues without the self-energy. They also agree with Fig.~6(b)
  of \Reference{Inotani2019}, where our $\rho_c/\rho_1$ corresponds to
  $N_{\text{fluct}}/N_{\text{MF}}$.}

This problem is solved in the other three methods by the subtraction
of the quasiparticle energy shift from the self-energy. However, as
mentioned in \Sec{subsec:occ-numbers-results}, some caution should be
used when comparing \Figs{densite_cut-off}(b-c) with the NSR result in
\Fig{densite_cut-off}(a), because the chemical potential $\mu$ does
not have the same meaning in the two cases. We see that among the
three approaches employing a subtraction, subtracting only the HF
self-energy [\Fig{densite_cut-off}(b)], as it was done
in~\cite{Ramanan2013}, yields a significantly larger correlated
density than the subtraction of the total self-energy
[\Fig{densite_cut-off}(c)] or the full Dyson equation
[\Fig{densite_cut-off}(d)]. It is interesting to notice that the last
two approximations give almost identical results for $\rho_c$ in spite
of the different shapes of the corresponding occupation numbers
[cf. \Fig{fig_nb_occupa_mu}].

Let us now discuss the results obtained with lower cutoffs $\Lambda =
1-1.8 \; \fmi$ (red, green and purple lines in
\Fig{densite_cut-off}). Keeping in mind the dependence of $T_{c}$ on
the cutoff, shown in \Fig{Tc-mu_cut-off}, we limit these curves to the
interval of $\mu$ where $T_c$ is cutoff independent. Nevertheless, we
notice that the correlated densities depend on the cutoff values, but
to a greater or lesser extent depending on the approximation that is
used. In the NSR case [\Fig{densite_cut-off}(a)], only the
$\Lambda=1\;\fmi$ result deviates significantly from the results
obtained with higher cutoffs. The strongest cutoff dependence, even
for $\Lambda=1.8\;\fmi$, can be seen in the calculation using the HF
subtraction [\Fig{densite_cut-off}(b)], while the cutoff dependence is
hardly visible for calculations using total self-energy subtraction
[\Fig{densite_cut-off}(c)] and the full Dyson equation
[\Fig{densite_cut-off}(d)]. This can be easily understood because the
HF self-energy $\Sigma_{\HF}$, \Eq{Sigma_HF}, uses directly the
interaction $v$ instead of the T matrix $\Gamma$, but the $V_{\lowk}$
interaction is made such that $\Gamma$ in vacuum is independent of the
cutoff, which implies a strong cutoff dependence of $v$ and hence of
$\Sigma_{\HF}$. Although the cutoff independence of $\Gamma$ in vacuum
does not necessarily ensure the cutoff independence of $\Gamma$ in the
medium, we see that the results that employ the full $\Gamma$, i.e.,
the subtraction of the total self-energy and the full Dyson equation,
satisfy the cutoff independence very well. It is, however, not clear
why for $\Lambda = 1 \; \fmi$, the cutoff dependence of the NSR result
sets in already at an unexpectedly small value of $\mu$.

%%%%%%%%%%%%%%%%%%%%%%%%%%%%%%%%%%%%%%%%%%%%%%%%%%%%%%%%%%%%%%%%%%%%%%%%%%%%%%%
\subsection{Density dependence of the critical temperature}
\begin{figure}
\begin{center}
\includegraphics[width=7.5cm]{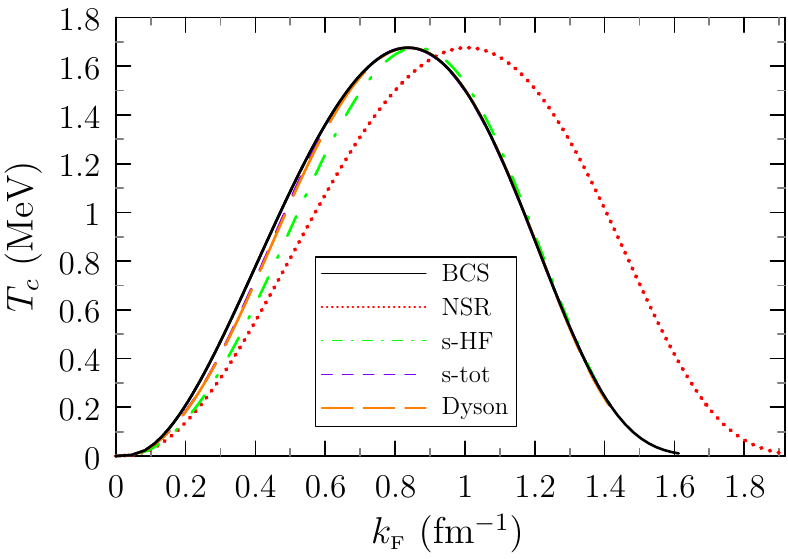}
\caption{Critical temperature $T_{c}$ as a function of $k_{\F}$
  calculated with the separable interaction with different
  approximations for the density: uncorrelated density (BCS, black
  solid line), NSR scheme (red dots), HF subtraction (green
  dashed-dotted curve), subtraction of the total self-energy (purple
  short dashes), full Dyson equation (orange long dashes).}
\label{graphe_temperature_critique}
\end{center}
\end{figure}

%%%%%%%%%%%%%%%%%%%%%%%%%%%%%%%%%%%%%%%%%%%%%%%%%%%%%%%%%%%%%%%%%%%%%%%%%%%%%%%
Having computed the critical temperature $T_c$ and the total density
$\rho$ as functions of the chemical potential $\mu$, we can now
compute $T_c$ as a function $\rho$. To emphasize the low-density
behavior, where strong-coupling effects are most pronounced, we show
in \Fig{graphe_temperature_critique} the critical temperature $T_c$
%%%%%%%%%%%%%%%%%%%%%%%%%%%%%%%%%%%%%%%%%%%%%%%%%%%%%%%%%%%%%%%%%%%%%%%%%%%%%%%
%%%%%%%%%%%%%%%%%%%%%%%%%%%%%%%%%%%%%%%%%%%%%%%%%%%%%%%%%%%%%%%%%%%%%%%%%%%%%%%%
as a function of the Fermi momentum $k_{\F}$, which is related to
$\rho$ by $k_{\F} = (3 \pi^{2}\rho)^{1/3}$.

Compared to the BCS result (black curve), which is computed with the
uncorrelated density $\rho_0$, the curves that account for the
correlations in the normal phase are more or less strongly shifted to
the right, depending on the correlated density $\rho_c$. The fact that
the NSR correlated density does not tend towards zero in the
high-density limit causes the NSR curve (red) to be shifted even at
the highest densities where $T_c$ is very low, as it was also found
in~\Reference{Inotani2019}. In the other treatments of the correlations, the
curves tend towards the BCS one at high density, which is the expected
result since this limit corresponds to the weak-coupling region where
the BCS theory should be valid. The correlated densities calculated
with the subtraction of the total self-energy (purple) and with the
full Dyson equation (orange) being almost identical, it is difficult
to distinguish these two curves in \Fig{graphe_temperature_critique}.

\begin{figure}
\begin{center}
  \includegraphics[width=7.5cm]{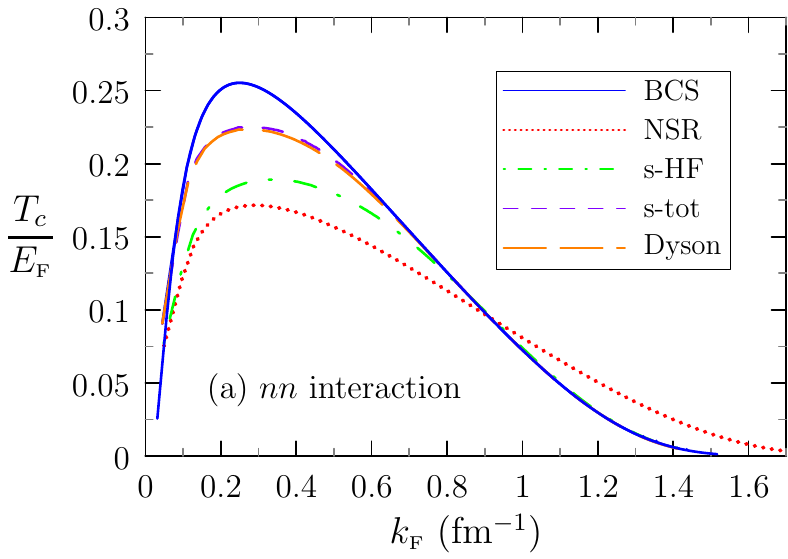}\hspace{4mm}
  \includegraphics[width=7.5cm]{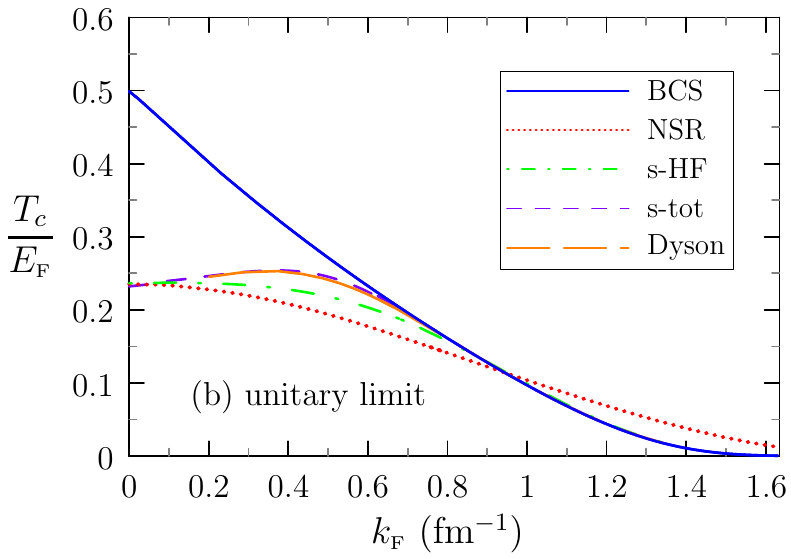}
\caption{Ratio $T_{c}/E_{\F}$ as a function of $ k_{\F} $ for the
  separable interaction. The blue solid curve represents the free case
  (BCS), the red dotted curve the NSR case, the green dashed-dotted
  curve the HF subtraction, the purple short dashed curve the 
  subtraction of the total self-energy, and the orange long dashed
  curve the density obtained with the full Dyson equation. 
  Panel (a) shows results for the physical $nn$
  interaction, while panel (b) shows results for the interaction that
  was readjusted to give an infinite $nn$ scattering length (unitary
  limit).}
\label{Tc-Ef_separable}
\end{center}
\end{figure}

In order to better see where the weak- and strong coupling regimes are
located, we plot in \Fig{Tc-Ef_separable}(a) the ratio $ T_{c}/E_{\F}$ as
a function of $k_{\F}$, where $E_{\F} = \hbar^2k_{\F}^2/(2m)$ denotes
the Fermi energy.
%%%%%%%%%%%%%%%%%%%%%%%%%%%%%%%%%%%%%%%%%%%%%%%%%%%%%%%%%%%%%%%%%%%%%%%%%%%%%%%
%%%%%%%%%%%%%%%%%%%%%%%%%%%%%%%%%%%%%%%%%%%%%%%%%%%%%%%%%%%%%%%%%%%%%%%%%%%%%%%
The BCS theory should hold in the weak-coupling limit $T_{c}/E_{\F}\ll
1$. Therefore, largest deviation from the BCS solid curve (blue)
should be seen around $k_{\F} \approx 0.3 \; \fmi$, where the ratio
$T_c/E_{\F}$ has its maximum. Again, as discussed above, the NSR
dotted curve (red) is the only one that does not tend towards the BCS
curve at high density. In the case of HF subtraction (green short
dashed curve), used in \Reference{Ramanan2013}, a significant reduction of
$T_c$ compared to the BCS one is seen in the interval $k_{\F} \approx
0.1 - 0.8\;\fmi$ ($\rho \approx 5\cdot 10^{-5} - 2\cdot
10^{-2}\;\fm^{-3}$), and the reduction can be up to $\sim
30\,\%$. With the subtraction of the total self-energy (purple short
dashed curve) or the full Dyson equation (orange long dashed curve), a
significant deviation from the BCS result is observed only in the
smaller interval $k_{\F} \approx 0.1 - 0.6\;\fmi$ (i.e., $\rho\sim
5\cdot 10^{-5} - 7\cdot 10^{-3}\;\fm^{-3}$) and the reduction is at
most $\sim 12\,\%$.

The small correlated densities that we get with the subtraction of the
total self-energy or with the Dyson equation are quite astonishing. In
ultracold atoms (i.e., with a contact interaction) in the unitary
limit, the experimental result is $T_c/E_{\F} \simeq 0.16$
\cite{Nascimbene2010,Ku2012}, while the BCS theory predicts
$T_c/E_{\F} \simeq 0.5$. A large part of the observed suppression is
believed to come from the correlated pairs above $T_c$, for instance
with NSR one finds $T_c/E_{\F}\simeq 0.23$ \cite{SadeMelo1993}, i.e.,
a suppression by more than $50\,\%$ compared to BCS. The remaining
reduction could be, e.g., due to screening effects \cite{Pisani2018},
but within the self-consistent Green's function formalism one finds
$T_c/E_{\F} \simeq 0.16$ even though screening is not included
\cite{Haussmann2007}.

Let us therefore test our methods by verifying that we reproduce the
known results. To that end, we repeat the calculations with the
modified value of the coupling constant $g$ of the separable
interaction corresponding to the unitary limit given in
Table~\ref{tab_fit}. Figure~\ref{Tc-Ef_separable}(b) shows us the
ratio $T_{c}/E_{\F}$ as a function of $k_{\F}$ for this case. We
notice that $T_{c}/E_{\F}$ remains finite even in the limit $k_{\F}\to
0$. This is a peculiarity of the unitary limit, because the
dimensionless parameter $k_{\F} a$ remains infinite independently of
the value of $k_{\F}$. However, the finite range $r_0$ of the
interaction becomes negligible in this limit because the relevant
dimensionless combination $k_Fr_0$ tends to zero. Hence, in the limit
$k_{\F} \to 0$, our results reproduce those obtained in ultracold
atoms with a contact interaction: In BCS, we find $T_{c}/E_{\F}\simeq
0.5 $ and in the correlated cases (NSR, HF subtraction, total
self-energy subtraction, and full Dyson equation) we find
$T_{c}/E_{\F}\simeq 0.23$ which is compatible with the results
obtained in~\cite{SadeMelo1993} for NSR and in~\cite{Pantel2014} for
the subtraction of the total self-energy (denoted there
Zimmermann-Stolz (ZS) scheme after \Reference{Zimmermann1985}).

For finite values of $k_{\F}$, the interaction gets weaker because of
its momentum dependence, and we expect that, as a function of $k_\F$,
we should find a similar behavior as if one increases the parameter
$-1/(k_{\F}a)$ in the BEC-BCS crossover of cold atoms. All curves
except the NSR one tend towards the BCS result. We also find that the
behavior of the curves for NSR and total self-energy subtraction is
similar to that obtained in cold atoms, namely that the NSR curve is
strictly increasing with increasing interaction on the BCS side (since
the maximum lies on the BEC side \cite{SadeMelo1993}), while the curve
for total self-energy subtraction has its maximum on the BCS side
\cite{Pantel2014}.

From this comparison we conclude that the smallness of the correlation
correction that we find in \Fig{Tc-Ef_separable}(a) is a consequence
of the finite scattering length $a = -15.9 \;\fm$
of the realistic separable interaction. Although this value is quite
large compared to other nuclear scales, the parameter $k_{\F} a$ gets
only large when the finite range of the interaction already starts to
weaken it, so that the BCS-BEC crossover effects remain rather weak.

%%%%%%%%%%%%%%%%%%%%%%%%%%%%%%%%%%%%%%%%%%%%%%%%%%%%%%%%%%%%%%%%%%%%%%%%%%%%%%%%
\section{Open questions}
\label{sec:open}
%%%%%%%%%%%%%%%%%%%%%%%%%%%%%%%%%%%%%%%%%%%%%%%%%%%%%%%%%%%%%%%%%%%%%%%%%%%%%%%%
\subsection{Problem of the subtraction}\label{subsec:problem-subtraction}
\begin{figure}
\begin{center}
\includegraphics[width=7.5cm]{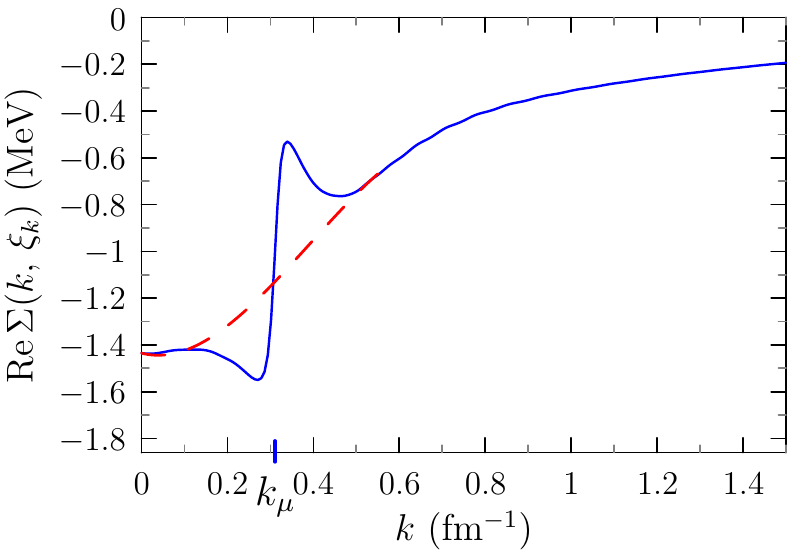}
\caption{Real part of the on-shell self-energy $ \Sigma(k,\xi_k)$ as a
  function of $k$ for $\mu = 2 \; \MeV $ and $T = 1.01 \, T_{c}(\mu) =
  0.53 \; \MeV$, computed with the separable interaction (blue
  curve). The value of $ k_{\mu} = \sqrt{2 m\smash{\mu}} / \hbar $
  approximates the position of the Fermi surface. The red dotted line
  is not the result of a calculation but it is drawn by hand to show
  schematically how a more appropriate subtraction could look like.}
\label{Uk_sep}
\end{center}
\end{figure}

%%%%%%%%%%%%%%%%%%%%%%%%%%%%%%%%%%%%%%%%%%%%%%%%%%%%%%%%%%%%%%%%%%%%%%%%%%%%%%%%
In \Sec{subsec:occ-numbers-formalism}, we argued that in a
non-self-consistent treatment of the propagators, we have to subtract
from the self-energy the mean-field like quantity $\Sigma(k,\xi_k)$.
In \Fig{Uk_sep}, we represent the momentum dependence of this shift in
the case of the separable interaction.
%%%%%%%%%%%%%%%%%%%%%%%%%%%%%%%%%%%%%%%%%%%%%%%%%%%%%%%%%%%%%%%%%%%%%%%%%%%%%%%%
%%%%%%%%%%%%%%%%%%%%%%%%%%%%%%%%%%%%%%%%%%%%%%%%%%%%%%%%%%%%%%%%%%%%%%%%%%%%%%%%
By defining $ k_{\mu} = \sqrt{2 m \smash{\mu}} / \hbar $ which roughly
indicates the position of the Fermi surface, we notice a big jump in
this region. It is clear that a momentum dependent mean field as
computed with, e.g., a Skyrme or Gogny interaction, would never have
such a shape. We suspect that by doing this subtraction, while
computing the internal propagators in $\Gamma$ and $\Sigma$ with the
free dispersion relation, we remove parts of the physical effects
caused by the self-energy. It seems probable that this subtraction, as
it was implicitly also used in
\cite{Zimmermann1985,Schmidt1990,Stein1995,Jin2010,Pantel2014},
reduces the pseudogap and the correlated densities. We notice that in
the literature there are other possible ways to do this
subtraction. One of them, used in \cite{Perali2002,Pieri2004},
consists in neglecting the momentum dependence and taking it at $k =
k_{\mu}$, i.e., subtracting a constant $\Sigma_0 = \Re \Sigma
(k_{\mu}, 0)$. Looking at the graph, we see that this is problematic,
too, because the subtraction is precisely fixed in the zone of the
jump where the function varies rapidly, so that the value of the
subtraction depends very sensitively on the approximation.
Qualitatively, we think that it would be more appropriate to subtract
a smooth curve as schematically represented in \Fig{Uk_sep} by the red
dotted line.

%%%%%%%%%%%%%%%%%%%%%%%%%%%%%%%%%%%%%%%%%%%%%%%%%%%%%%%%%%%%%%%%%%%%%%%%%%%%%%%%
\subsection{Effect of the quasiparticle weight}\label{subsec:Z-factor}
%%%%%%%%%%%%%%%%%%%%%%%%%%%%%%%%%%%%%%%%%%%%%%%%%%%%%%%%%%%%%%%%%%%%%%%%%%%%%%%%
In \Reference{Cao2006}, a new effect was discussed that strongly reduces the
critical temperature, namely the reduced quasiparticle weight. It is
well known that the Cooper instability comes from the jump of the
Pauli factor $1-\ff(\xi_{\kk/2+\pp})-\ff(\xi_{\kk/2-\pp})$ in the
numerator of \Eq{G02} at $p=k_{\mu}$ for $k=0$. However, this Pauli
factor is obtained with uncorrelated Green's functions having a
quasiparticle weight of unity, $Z=1$. In \Reference{Cao2006}, it was argued
that the amplitude of the jump should be multiplied by the
quasiparticle weight $Z < 1$, which in that work was determined from a
Br\"uckner calculation.

A more self-consistent method to include this effect would be the
so-called renormalized RPA (r-RPA) \cite{Schuck2020}, where, in the
particle-particle channel, the Pauli factor is replaced by
\begin{equation}\label{r-RPA}
  1-\ff(\xi_{\kk/2+\pp})-\ff(\xi_{\kk/2-\pp}) \to
    1-n(\kk/2+\pp)-n(\kk/2-\pp)\,,
\end{equation}
with $n(k)$ the correlated occupation numbers, which should be
calculated self-consistently. At zero-temperature, the correlated
occupation numbers have a jump of height $Z$ at $k=k_{\F}$ and
therefore the height of the jump of the Pauli factor is also reduced
by a factor of $Z$. We recently applied this method to the case of
strongly polarized Fermi gases at zero temperature \cite{Durel2020}
where it reduces the critical polarization, which plays a similar role
as the critical temperature in the present case.

However, here it turns out that it is not possible to realize a
self-consistent calculation of r-RPA type. When we insert the
correlated occupation numbers in the Pauli factor $\bar{Q}(k,p)$, it
does no longer vanish at $p = k_\mu$, leading to a non-vanishing
imaginary part of the vertex function $\Gamma$ at $\omega = 0$. This
makes it impossible to calculate the self-energy, as one can see from
\Eq{ImSigma}: $\Im\Gamma$ must vanish where the Bose function $\fb$
diverges.

Despite this problem, let us estimate the order of magnitude of the
effect. To that end, we replace in the angle averaged Pauli factor
$\bar{Q}$ the free occupation numbers by the correlated ones according
to \Eq{r-RPA}, calculated with the full Dyson equation with the
separable interaction. For this choice, in certain regions (in the not
too strongly coupled regime), the imaginary part of $\Im J(k=0,\omega)$
[cf.~\Eq{def-J}] changes sign almost at $\omega=0$ so that we can try
to estimate the critical temperature from $\Re J(k=0, \, \omega=0 \, ; \,T=T_c)
= 1/g$ [cf.~\Eq{Gamma-separable}].

\begin{figure}
\begin{center}
\includegraphics[width=7.5cm]{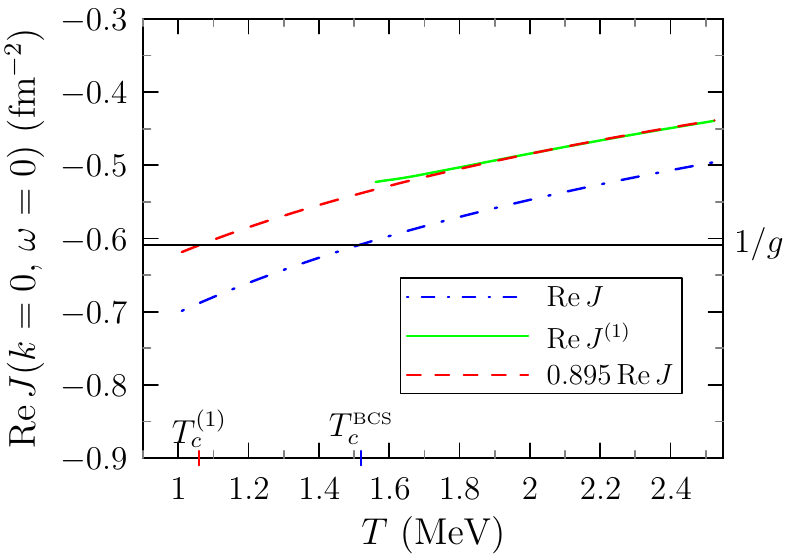}
\caption{Real part of $ J(k=0, \, \omega=0)$ for $\mu = 20 $ MeV as a
  function of the temperature $T$. The blue dash-dot curve represents
  the usual calculation, using $\bar{Q}$ according to \Eq{def-Qbar},
  while the the green curve displays $J^{(1)}$ which is obtained when
  the Fermi functions in the Pauli factor are replaced according to
  \Eq{r-RPA}. The red dashed curve is an extrapolation of $ J^{(1)} $
  down to $T_c^{(1)}$ under the assumption that it has the same form
  as $ J $.}
\label{fig_auto-coherence}
\end{center}
\end{figure}

This is illustrated in~\Fig{fig_auto-coherence}.
%%%%%%%%%%%%%%%%%%%%%%%%%%%%%%%%%%%%%%%%%%%%%%%%%%%%%%%%%%%%%%%%%%%%%%%%%%%%%%%
%%%%%%%%%%%%%%%%%%%%%%%%%%%%%%%%%%%%%%%%%%%%%%%%%%%%%%%%%%%%%%%%%%%%%%%%%%%%%%%
Notice that the occupation numbers $n(k)$ used in \Eq{r-RPA} are
computed with the ordinary $J$ and not self-consistently, because, as
mentioned above, we cannot compute the self-energy with the modified
$J$. Hence, our modified $J$, denoted $J^{(1)}$, can be regarded as a
first step of a self-consistent iteration. Furthermore, as it is
impossible to compute the correlated occupation numbers below
$T_c^{\BCS}$, it is necessary to extrapolate $ J^{(1)} $ to the new
critical temperature $T_c^{(1)}$. In the example $\mu = 20 $ MeV
presented in \Fig{fig_auto-coherence}, we find that at temperatures
not too close to $T_c^{\BCS}$ this effect reduces $\Re J(0,0)$ by
about $10\,\%$, reducing $T_c$ by about $30\,\%$. Although this is
only a rough estimate, it indicates that this effect might be
important.

%%%%%%%%%%%%%%%%%%%%%%%%%%%%%%%%%%%%%%%%%%%%%%%%%%%%%%%%%%%%%%%%%%%%%%%%%%%%%%%%
\section{Conclusions}
\label{sec:conclusions}
%%%%%%%%%%%%%%%%%%%%%%%%%%%%%%%%%%%%%%%%%%%%%%%%%%%%%%%%%%%%%%%%%%%%%%%%%%%%%%%%
The in-medium T-matrix formalism makes it possible to describe pair
correlations in neutron matter. To model the $ nn $ interaction, we
use the effective low-momentum interaction $V_{\lowk}$. We presented a
numerical method to solve the integral equation for the vertex
function $\Gamma$. It is also possible to replace $V_{\lowk}$ by a
separable interaction which gives very similar results and largely
facilitates the numerical computations.

Because of the large $nn$ scattering length, it is believed that
dilute neutron matter is close to the unitary Fermi gas and that
BCS-BEC crossover physics plays an important role. We pointed out that
the standard NSR
theory, which is widely used for the BCS-BEC crossover, has the
shortcoming not to tend towards the weak-coupling BCS theory at high
density. It leads to unphysical occupation numbers and to an incorrect
shift of the $T_c$ vs. $\rho$ curve, because it attributes the
mean-field like shift of the quasiparticle energy to the
correlations.

We compared different prescriptions to avoid these deficiencies by
subtracting this shift from the dynamical (i.e., energy dependent)
part of the self-energy. Subtracting only the Hartree-Fock (HF) part,
as in \Reference{Ramanan2013}, is already enough to make the correlations
tend towards zero in the weak-coupling limit. To go beyond this
approximation, we subtract the total on-shell self-energy, as
implicitly done within the Zimmermann-Stolz scheme
\cite{Zimmermann1985} which was previously used for the description of
the BCS-BEC crossover in dilute symmetric nuclear matter
\cite{Schmidt1990,Stein1995,Jin2010}. This prescription avoids the
unphysical dependence of the HF subtraction on the cutoff of the
$V_{\lowk}$ interaction, but has the effect of further reducing the
correlations and therefore the crossover effects. With the explicit
calculation of the self-energy, we are also able to calculate the
correlations through the full Dyson equation and not just its
truncation to the first order as it is customary to do. This method
gives us access to the spectral function and allows us to
  highlight the existence of a pseudo-gap in neutron matter above
$T_c$. As discussed in \Sec{subsec:problem-subtraction}, it is
possible that we underestimate the pseudogap as a consequence of the
subtraction method.

At a quantitative level, it seems that the suppression of the critical
temperature for a given density, due to the density of preformed
pairs, is too weak to have much effect on neutron-star
observables. However, a more important effect is expected to result
from the quasiparticle weight $Z<1$, as also pointed out in
\Reference{Cao2006}. In the weak-coupling regime we estimate that this
effect may reduce $T_c$ by $\sim 30\,\%$, but our present theory does
not allow us to compute it self-consistently or in the strong-coupling
regime.

Finally, a large uncertainty comes from screening of the bare $nn$
interaction in the medium \cite{Cao2006,Ramanan2018,Ramanan2020} which
is not included in the present study. These effects depend sensitively
on the approximations that are used, on the Landau parameters and on
the momentum dependence of the effective interaction used in the
residual particle-hole interaction. From the results of
\Reference{Ramanan2020}, it seems most likely that in the low-density region
corresponding to the strong-coupling regime, screening suppresses the
critical temperature by about $30-40\,\%$, thereby reducing further
the crossover effects.

\bibliography{bibliographie}

\end{document}